\documentclass[11pt]{article}
\usepackage{graphicx}
 \usepackage{color}
\usepackage{ulem}
\usepackage{epsfig}
\usepackage{latexsym}
\usepackage{amsmath}

\begin{document}

\title{\Large \bf   
  New inequality   indicators for  team ranking in multi-stage female professional cyclist races}
    \bigskip  
 \author{ \large \bf  Marcel Ausloos \textsuperscript{1,2,3,4}   \\\\ \\
$^1$ School of Business,
University of Leicester,    Brookfield, \\ Leicester, LE2 1RQ, UK    \\$e$-$mail$ $address$:
ma683@leicester.ac.uk \\ \\
$^2$ Group of Researchers Applying Physics in Economy and Sociology \\(GRAPES),  Beauvallon, rue de la Belle Jardini\`ere, 483/0021\\ Sart Tilman,  Angleur, B-4031, Li\`ege, Belgium  \\
  e-mail: marcel.ausloos@uliege.be\\ \\
$^3$ Department of Statistics and Econometrics, \\The Bucharest University of Economic Studies,  \\ Caeia Dorobantilor 15-17,  010552 Bucharest, Romania \\$e$-$mail$ $address$:marcel.ausloos@csie.ase.ro \\ \\
$^4$ Department of Statistics, Predictions and Mathematics, \\Universitatea Babe{\c{s}}-Bolyai, \\Str. Mihail Kogălniceanu 1, 400084, Cluj-Napoca, Romania   \\ \\\\}

\maketitle
\newpage
  \begin{abstract}
Cycling competition is highly interesting since the team ranking is based on the best performance of some subset of team members.  The paper develops new inequality indicators,  a methodology to construct them, and numerical illustrations allowing to provide operative arguments in their favor.  The numerical illustrations subsequently deal with hierarchical ranking indicators of (female) cyclist teams, competing in multi-stage races.   For the illustration, 
  the 2023 editions of the most famous  long races for females are considered:  
 34th Giro d'Italia Donne, 2nd  Tour de France Femmes, 9th Vuelta Femenina.
 
 Several classical ranking indicators   are recalled and adapted to the study cases.   
The most usual indicator,   $T_L$,  is  based on the riders arriving time for the various stages, i.e., according to  Union Cycliste Internationale (UCI) standard rules.  One also uses another indicator,  $A_L$,  which requires that the riders finish  the race, whence each stage, in order to define the race best team.
Another contribution of the paper derives from specific developments of these indicators, thereby leading to  new measures:   the   “leadership gap" based on $A_L-T_L$, and  the “competition temperature", based on entropy. It is argued that the numerical values point to differences in team strategy based on rider skill levels.  The ranking of contributions to indicators allow to observe the  "crucial core"  made of the  most competitive teams.

 \end{abstract}
 
  
\newpage
 
\section{Introduction} \label{Introduction}
 
 Rules have to be devised for providing a realistic hierarchy of choices. 
 
   Yet,   the ranking methodology     can lead to much debate,$^{\cite{Kulakowski20,Fritzetal23,AusloosGRRC24,Kossi17,ref3academiachoice,Jose,CsatoIGTR2019} }$ 
 among many others in social choice considerations, as  in the case of tournament ranking methods. 
$^{\cite{Kulakowski20,ChebotarevShamis1998,Fainmesseretal2009,GonzalezDiazetal2014,Csato2017,Vazirietal2018,Csato2023,LeivaBertran2025}}$
In particular,  sport activities seem to provide rather objective and quantitatively reliable data  for academic studies.$^{\cite{JianuAoR2023}}$

Among these, it appears that cyclist races contain much interesting data. Indeed, one can focus on the role of individuals within a team, 
 since team hierarchy is based on the best performance of a subset of members of the team.
 Within this framework, incentive must be provided to teams and team members for showing some interesting race competition.$^{\cite{CsatoEJOR2022}}$ 
Thus, relevant  hierarchy indices are needed, - and somewhat tied to money awards.$^{\cite{ausloos2017hintTdF,ausloos2020concentrationTdF}}$
   
In the following, one considers   the three most famous female races within the Union Cycliste Internationale (UCI) classifications: Giro d'Italia Donne,
 Tour de France Femmes,
Vuelta Femenina.

According to UCI rules,  the hierarchy of the teams,  at the end of an $L$-long multi-stage (professional) cyclist race, depends on the cumulative time ( $T_L$) of the team's 3 fastest riders  for every stage, - not taking into account time bonuses or penalties. These   times are  relevant for the rider standing, but should not be taken into account   for the team rank.
However, for the final team ranking at the end of the race, UCI team hierarchy does not even care if such riders, relevant for some stages, do finish the  whole race. This highly debatable measure has been discussed elsewhere as paradoxical - and proved to be highly  biased.

Thus, 
one may introduce an  “adjusted team final time"  measure,  $A_L$,  based on the (3 fastest)  riders of a team who have finished the whole race.$^{\cite{Ausloos2023,EJOR}}$   In so doing, one avoids possible Cipollini effect,$^{\cite{ANORCipollini}}$   -  when riders are specifically  selected for the few first usually easy stages, as sprinters, but are withdrawn thereafter, yet globally contributing, even though absent, to the overall team time classification.

  Going beyond the above consideration,  one may derive metrics that aim at  measuring some team skill and  also at attempting to quantify team global strategy, for a  given race. For so doing, one proposes two new measures or indicators: (i)  the  “leadership gap index",  (ii) the  “race temperature  index". The ranking of teams according to such indicators allow to observe the  “crucial core"  made of the  most competitive teams.

In brief, these two so newly defined metrics complement the entropy approach$^{\cite{Ausloos2023}}$ and hopefully develop previous works$^{\cite{EJOR,ANORCipollini}}$ toward team management and coaching applications.
 
  For completeness, let it be observed that this paper enters the framework of studies on cycling published in  the International
Journal of Sports Science \& Coaching and other journals.  Notice that most works pertain to the (physiological) characteristics of the cyclists.$^{\cite{Mostaertetal2021,PinedojauregiRomarate2025,Smith2008,Stessesetal2024}}$ 
 Closer to the present aim,  O'Grady et al. discuss, after interviews, tactical strategies that professional riders and coaches prepare at training time for application in races.$^{\cite{OGradyetal2023}}$
 
In Section \ref{RQ}, one  poses the Research Questions and  mentions the Data sources.  
One displays the fundamental characteristics of the  races.

In Section \ref{Methodology}, one introduces the methodology, including the  formulae for $T_L$ and $A_L$. 
 Next,  one explains that (i)  the  “leadership gap index",  in Section \ref{leadershipgap}, is based on   [$A_L-T_L$];  (ii) the  “race temperature  index" is defined through  Shannon-Boltzmann-Gibbs entropy concepts, in Section \ref{ShannonStageTemperature}. 
 It is argued why these indices are so called.  
 
 In Section \ref{Others}, other hierarchy measuring indices are considered   for readily comparison, i.e., some qualitative advantages and disadvantages of the newly proposed indicators.$^{\cite{Vazirietal2018,Sokolov2012,Subramanian2012,ZIDMA,JosaAguado2020,Bednayetal25}}$
   There exist  fundamentally different approaches in ranking methodology.    It is pertinently emphasised that changing the ranking rules,  in a multi-stage race, may change some tournament metrics; see, for example, scoring and ranking simulation by    Csató.$^{\cite{Csato2023}}$  
   
  Here, two    inequality  indicators can be  directly derived from the distribution characteristics in order to evaluate the dispersion of team “values": the Atkinson index,$^{\cite{Atkinson1970}}$
 and the Coefficient of Variation. Three classical indicators of dispersion can be next considered: the Herfindahl-Hirschman index,  the Gini coefficient, and the Theil index.$^{\cite{Hirschman,Gini1921,Theil65}}$ These indicators  
show how dispersed the final times are,  but are calculated without taking into account the ranking of teams.  One may  delve more into the hierarchy problem if one ranks the components. Moreover, one can calculate other {\it ad hoc} indicators: the Pietra-Hoover index,$^{\cite{MagesRohner2024}}$ and the Rosenbluth coefficient.$^{\cite{HallTideman67}}$

 Section \ref{ResultsandAnalysis} contains numerical results and some analysis.  In Section \ref{Discussion}, one deepens into  team hierarchy, comparing  teams  in the various races.   Conclusions follow in Section \ref{conclusions}, together with suggestions for further research due to obvious limitations of the present study.

  \section{Research Questions} \label{RQ}

  Due to the considerations outlined in the Introduction section, i.e.,  UCI unjustified shocking biased constraints on usual team value measures, one can select  the following research questions as a guiding thread of the paper:  
  \begin{itemize}
  \item can one provide indicators with less biased constraints  on team ranking?
  \item  are they  strategic or coaching   
  features which arise in studying and measuring team “competition" hierarchy, in cyclist races? 
       \end{itemize}
   For finding proper answers geared toward various disciplines but based on case studies the following top multistage races with complex data, are hereby used:
  \begin{itemize}
  \item
34th Giro d'Italia Donne;
  \item
 2nd Tour de France Femmes;
  \item
9th Vuelta Femenina.
    \end{itemize}
    For the present exposé only a single year is examined: 2023.

  For simplicity, the races will be called $TdF$, $GdI$, $VaE$.      A few fundamental characteristics for the 3 races  are found in Table \ref{TableTdFGdVaEILn0M0_MandF}:  dates of races, length, number of stages, number of riders and of teams, etc. 
  The list of competing teams and their UCI code are given in Table \ref{TableUCIacronMandF}, also emphasizing the team level according to UCI for those participating in a specific race: $W$ refers to Women's World Teams; $C$ to UCI Women's Continental Teams.

  One can freely obtain relevant data from the organizers websites. However, they are not all provided in a consistent way. Therefore, it is best to rely 
  on professional websites, e.g.,
  $ https://www.procyclingstats.com$. Nevertheless, data cross-checking  must be systematically done; one chosen method has been to use    $Wikipedia$ pages.  Disagreements are still found; they have been manually resolved.

 \begin{table}\begin{center}  \fontsize{10pt}{10pt}\selectfont
\begin{tabular}{|c||c|c|c|} 
 \hline  notations 		&   $GdI$ &   $TdF$ & $VaE$  \\  	
  \hline	\hline			
$L$			&  8(*)		& 8	& 7	 	\\	 
$n_0$		&7 		& 7 		& 7 		\\	 	
$M_0$		& 24 	(=15+9)	& 22 (=15+7) 		&23 (=12+11)		\\ 
distance ($km$) & 963.6 		& 960.4 	& 740.5 \\ \hline
$N_0$ 		&167 	& 154 	& 160	\\	 
$N_L$ 		&133 	& 123 	& 127 	\\	 
$M_L$		& 23 		& 22 		& 22		\\ \hline
winning rider 	& vanVleuten & Vollering & vanVleuten\\		
winning team	& MOV 	& SDW 	& UAD 	\\  \hline 
race dates  & 06/30-07/09  & 07/23-07/30 & 05/01-05/07\\ \hline 
	\end{tabular}
\caption{Characteristics values of the 2023 $GdI$, $TdF$, $VaE$ races;  $L$: the number of stages; $n_0$: the allowed maximum number of  starting riders in a team; $M_0$: the number of teams considered (distinguishing the number of  W and C teams);
  $N_0$ and $N_L$: the number of starting and finishing riders; $M_L$ is the number of teams  finishing the race with at least 3 riders.  The rider winner and the team winner (along UCI rules) and the race dates are recalled.
 (*) N.B. The race was supposed to be 9 stages long, i.e., 968 km, but the 4.4 km long (time trial) 1st stage was cancelled due to weather conditions.} 
\label{TableTdFGdVaEILn0M0_MandF}
 \end{center}
 \end{table}
  
 \begin{table}\begin{center}   \fontsize{9pt}{9pt}\selectfont
\begin{tabular}{|c||c|c||c|c|c|} \hline 	
 2022 &  UCI	& Team	&2023& 2023&2023\\ 
rank	&	code	&	Sponsors	&	$GdI$& $TdF$&$VaE$	\\\hline
		\hline
1	&	SDW		&	Team SD Worx		&	W	&	W	&	W	\\
2$^{(*)}$	&TFS	&	Trek - Segafredo // Lidl - Trek		&	W	&	W	&	W	\\
3	&	DSM	&	Team DSM - Firmenich	&	W	&	W	&	W	\\
4	&	FST	&	FDJ - SUEZ			&	W	&	W	&	W	\\
5	&	MOV	&	Movistar Team			&	W	&	W	&	W	\\
6	&	CSR	&	Canyon-SRAM Racing	&	W	&	W	&	W	\\
7	&	UAD	&	UAE Team ADQ		&	W	&	W	&	W	\\
9	&	JAY	&	Team Jayco AlUla		&	W	&	W	&	W	\\
10	&	JVW	&	Team Jumbo-Visma (TJV)	&	W	&	W	&	W	\\
11	&	TIB	&	EF Education-TIBCO-SVB&	W	&	W	&	W	\\
13	&	LIV	&	Liv Racing TeqFind		&	W	&	W	&	W	\\
14	&	WNT	&	CERATIZIT-WNT Pro Cycling&		&	C	&		\\
15	&	LPW	&	Lifeplus Wahoo			&		&	C	&		\\
19	&	AGS	&	AG Insurance - Soudal Quick-Step&	C	&	C	&		\\
20	&	HPU	&	Team Coop - Hitec Products&		&	C	&	C	\\
23	&	AUB	&	St Michel - Mavic - Auber93 WE&		&	C	&	C	\\
24	&	UXT	&	Uno-X Pro Cycling Team	&	W	&	W	&		\\
25	&	HPH	&	Human Powered Health	&	W	&	W	&		\\
26	&	BPK	&	BePink - GOLD			&	C	&		&	C	\\
28	&	COF	&	Cofidis Women Team	&		&	C	&		\\
30	&	ARK	&	Arkéa Pro Cycling Team	&		&	C	&		\\
31	&	MAT	&	Massi - Tactic Women Team&		&		&	C	\\
33	&	BDU	&	Bizkaia Durango&	C	&		&	C	\\
35	&	TOP	&	Top Girls Fassa Bortolo	&	C	&		&		\\
36	&	EIC	&	Eneicat - CMTeam - Seguros Deportivos&		&		&	C	\\
39	&	SWT	&	Sopela Women's Team	&		&		&	C	\\
40	&	VAI	&	Aromitalia - Basso Bikes - Vaiano&	C	&		&		\\
42	&	FBW	&	Farto-BTC Women's Cycling Team&		&		&	C	\\
51	&	LKF	&	Laboral Kutxa Fundación Euskadi&		&		&	C	\\
53	&	SBT	&	Isolmant - Premac - Vittoria&	C	&		&		\\
56	&	STC	&	Soltec Team			&		&		&	C	\\
59	&	BTW	&	Born To Win G20 Ambedo&	C	&		&		\\
63	&	CDR	&	Cantabria Deporte - Rio Miera&		&		&	C	\\
64	&	MDS	&	Team Mendelspeck		&	C	&		&		\\
112	&	FED	&	Fenix-Deceuninck		&	W	&	W	&		\\
145	&	COG	&	Israel Premier Tech Roland&	W	&	W	&	W	\\
$^{(**)}$&	GBJ	&	GB Junior Team Piemonte Pedale Castanese A.S.D.&	C	&		&		\\
		\hline			\end{tabular}
\caption{ UCI  acronym of team sponsors, for competing   female  teams in the 3 here considered  main multi-stage 2023 races ($GdI$, $TdF$, $VaE$); the  2022 ranking order is from $https://www.procyclingstats.com/rankings.php?date=2023-12-26\&nation=\&level=\&filter=Filter\&p=we\&s=uci-teams$. The 2022 UCI code acronyms are used. The various team participations into the specific races are summarized in the last 3 columns; W refers to Women's World Teams; C to UCI Women's Continental Teams, according to UCI 2023 levels. N.B.  $^{(*)}$ change of sponsors during racing year; $^{(**)}$ had no licence in 2022.
} \label{TableUCIacronMandF}
 \end{center}
 \end{table}
 
 \section{Methodology} \label{Methodology}

One should recall that in a $L$-long multi-stage cyclist race, the UCI rules imply that the winning team is discovered as the team  $(\#)$ having the lowest sum of the cumulative times for the 3 fastest riders of the team for each stage, i.e., the lowest $T_L$  
\begin{equation}\label{Tteameq}
  T_L^{(\#)}  = \Sigma_{s=1}^{L} \;\; t_s^{(\#)}  \;,
\end{equation}
when the   team (finishing) time  for    stage $s$ is 
 \begin{equation}\label{tsteameq}
t_s^{(\#)}  = \Sigma_{i=1}^{3} \;\;  t^{(\#)}_{i,s}\;,
\end{equation}
where $ t^{(\#)}_{i,s}$ is the finishing time of  one of the 3 fastest riders ($i=1,2,3$)  of  team $(\#)$ for that stage.

 It is re-emphasized that the fact that such riders do not necessarily  finish the $L$-long race appears to be irrelevant for UCI. But, one may rightly wonder thereafter whether the sum in Eq.(\ref{Tteameq}) points out to the “best team". 
 It seems that  one  should consider the  adjusted team final time $A_L^{(\#)}$ such that
 \begin{equation}\label{Ateameq}
  A_L^{(\#)}  = \Sigma_{j=1}^{3} \; \; t_{j,L}^{(\#)}
\end{equation}  
where $j = 1, 2, 3$ refers to the   3  fastest  riders   {\it having  completed all}  $L$ stages for team ${(\#)}$.$^{\cite{EJOR}}$ 
 Let it be  emphasized that these 3   “$j$" riders  {\it might} be quite different from the  3   “$i$" riders having contributed to any  $t_s^{(\#)}$, 
whence to   $T_L^{(\#)}$.  

 One complexity has to be emphasized: according to UCI rules,  the final time of a $rider$, at the end of a stage, whence at the end of the race, takes into account bonuses (and penalties); thus  the truly finishing time  of a rider is  apparently equal to the sum  of “the reported final time + bonus - penalties".  However, UCI rules disregard such extra time measures in order to calculate the $team$ time on a stage, whence for the whole race.  
 Thus, the same “restriction" has been used for calculating the $A_L$  team final time.
 
Therefore, the   methodological path goes as follows:  
\begin{itemize}
\item get each team $T_L^{(\#)} $, according to organizers published data.
\item  rank riders  in each team according to their true finishing race time, i.e., excluding  bonuses and 
  penalties (if they exist);
\item  select the 3 fastest riders overall in each team, for each daily stage, and add their  cumulative times to get $A_L$.
\end{itemize}

One obtains the values and hierarchy displayed in Tables \ref{Table4TLALGdI23teams} - \ref{Table5TLALVaE23teams}, 
in increasing time order.
 The statistical characteristics of the relevant finishing times distributions are reported in Table  \ref{TableTLALDL23Fteamtimescharacteristics}.
 
   \begin{table}\begin{center}
\begin{tabular}{|c||c|c|c||c|c|c|c||c|c|c|c|} 
 \hline 
&\multicolumn{6}{|c|}{$GdI$ 2023}   \\	\hline
 \hline
rank&team 	& $T_L^{(\#)}$ &  team 		&  $ A_L^{(\#)} $  &team& $\Delta_L^{(\#)}$    \\ \hline \hline   
1	&	MOV      	&	73:54:51	&	    MOV      	&	73:59:03	&	     DFP  	&	0:00:33	\\		
2	&	FST      	&	73:55:37	&	    FST      	&	74:02:34	&	    SDW   	&	0:03:13	\\		
3	&	LTK    	&	74:05:36	&	    DFP  	&	74:16:58	&	     VAI    	&	0:03:27	\\		
4	&	DFP  	&	74:16:25	&	   SDW   	&	74:29:31	&	     BPK   	&	0:04:01	\\		
5	&	CSR   	&	74:20:26	&	    UAD      &	74:35:06	&	     MOV    &	0:04:12	\\				
6	&	UAD      	&	74:22:33	&	    LTK    	&	74:35:41	&	     TJV   	&	0:05:45	\\		
7	&	SDW   	&	74:26:18	&	    CSR   	&	74:43:13	&	     FST     &	0:06:57	\\			
8	&	JAY   	&	74:32:19	&	    FED     	&	74:56:18	&	     COG    &	0:07:46	\\			
9	&	FED     	&	74:39:03	&	    TIB    	&	75:02:57	&	     TOP    	&	0:08:20	\\		
10	&	TIB    	&	74:43:13	&	    JAY   	&	75:05:57	&	     HPH    &	0:08:46	\\			
11	&	TJV   	&	75:04:06	&	    TJV   	&	75:09:51	&	     GBJ     &	0:09:06	\\			
12	&	HPH     	&	75:08:40	&	    HPH     	&	75:17:26	&	     UAD    &	0:12:33	\\			
13	&	AGS     	&	75:22:19	&	    COG    	&	75:38:24	&	     SBT     &	0:14:35	\\			
14	&	LIV    	&	75:23:08	&	    AGS     	&	75:39:49	&	     FED     &	0:17:15	\\			
15	&	COG    	&	75:30:38	&	    LIV    	&	75:45:24	&	     AGS     &	0:17:30	\\			
16	&	UXT     	&	75:34:50	&	    UXT     	&	75:57:22	&	     TIB    	&	0:19:44	\\		
17	&	TOP    	&	76:54:08	&	    TOP    	&	77:02:28	&	     MDS    &	0:21:56	\\			
18	&	BPK   	&	77:01:00	&	    BPK   	&	77:05:01	&	     LIV    	&	0:22:16	\\		
19	&	MDS    	&	77:02:05	&	    MDS    	&	77:24:01	&	     UXT     &	0:22:32	\\			
20	&	SBT     	&	77:49:13	&	    SBT     	&	78:03:48	&	     CSR   	&	0:22:47	\\		
21	&	VAI    	&	78:20:07	&	    VAI    	&	78:23:34	&	     LTK    	&	0:30:05	\\		
22	&	BDU      	&	78:22:49	&	    GBJ     	&	79:18:17	&	     JAY   	&	0:33:38	\\		
23	&	GBJ     	&	79:09:11	&	    BDU      &	79:30:07	&	     BDU    &	1:07:18	\\				
24	&	BTW  	&	...		&	    BTW   	&	...		&	      BTW  	&	...	\\\hline				
 \end{tabular}
\caption{Table displaying the  final ranking of  the  ($M_L=23$ Female) teams after the last $GdI$ 2023  stage, i.e., $T_L$ according to the UCI rules, or $A_L$  resulting from the sum of the   finishing times  of the 3 fastest riders (excluding their possible bonuses and penalties) of the team ($\#$);  the last two columns report the team hierarchy according to the ascending value of  $\Delta_L^{(\#)}\equiv A_L ^{(\#)}- T_L^{(\#)}$.  BTW finished the multi-stage race with only 2 riders.
 } \label{Table4TLALGdI23teams}.   
 \end{center}
 \end{table}

   \begin{table}\begin{center}
\begin{tabular}{|c||c|c|c||c|c|c|c||c|c|c|c|} 
 \hline 
&\multicolumn{6}{|c|}{$TdF$ 2023}   \\	\hline
 \hline
  rank& team 		& $T_L^{(\#)}$ &  team 		&  $ A_L^{(\#)} $  &team& $\Delta_L^{(\#)}$    \\ \hline \hline    
1	&	 SDW   	&	76:17:38	&	   SDW   	&	76:26:34	&	ARK   	&	0:02:12	\\
2	&	     CSR   	&	76:29:43	&	    MOV      	&	76:41:25	&	UAD      	&	0:04:15	\\
3	&	     MOV      	&	76:35:41	&	    CSR   	&	76:46:20	&	JAY   	&	0:05:34	\\
4	&	     FST      	&	76:37:18	&	    UAD      	&	76:53:19	&	COF     	&	0:05:35	\\
5	&	     UAD      	&	76:49:04	&	    FST      	&	76:54:55	&	COG    	&	0:05:37	\\
6	&	     AGS     	&	76:53:31	&	    AGS     	&	77:01:21	&	MOV      	&	0:05:44	\\
7	&	     TJV   	&	77:02:04	&	    TJV   	&	77:18:35	&	HPH     	&	0:05:53	\\
8	&	     LTK    	&	77:05:09	&	    COG    	&	77:20:44	&	AGS     	&	0:07:50	\\
9	&	     DFP  	&	77:07:58	&	    DFP  	&	77:24:11	&	WNT     	&	0:08:26	\\
10	&	     COG    	&	77:15:07	&	    LTK    	&	77:33:18	&	SDW   	&	0:08:56	\\
11	&	     FED     	&	77:27:11	&	    WNT     	&	77:37:40	&	AUB      	&	0:13:51	\\
12	&	     LIV    	&	77:27:47	&	    AUB      	&	77:55:05	&	DFP  	&	0:16:13	\\
13	&	     WNT     	&	77:29:14	&	    FED     	&	77:56:01	&	TJV   	&	0:16:31	\\
14	&	     TIB    	&	77:39:34	&	     COF     	&	77:57:04	&	CSR   	&	0:16:37	\\
15	&	     AUB      	&	77:41:14	&	    HPH     	&	78:00:33	&	FST      	&	0:17:37	\\
16	&	      COF     	&	77:51:29	&	    JAY   	&	78:06:15	&	LPW     	&	0:20:49	\\
17	&	     HPH     	&	77:54:40	&	    ARK   	&	78:17:01	&	LTK    	&	0:28:09	\\
18	&	     LPW     	&	77:58:42	&	    LIV    	&	78:17:50	&	FED     	&	0:28:50	\\
19	&	     JAY   	&	78:00:41	&	    LPW     	&	78:19:31	&	HPU    	&	0:38:39	\\
20	&	     UXT     	&	78:08:20	&	    TIB    	&	78:32:07	&	LIV    	&	0:50:03	\\
21	&	     ARK   	&	78:14:49	&	    UXT     	&	78:58:52	&	UXT     	&	0:50:32	\\
22	&	     HPU    	&	79:25:34	&	    HPU    	&	80:04:13	&	TIB    	&	0:52:33	\\
     \hline
 \end{tabular}
\caption{Table displaying the  final ranking of  the $M_L=22$   teams after the last $TdF$ 2023  stage, i.e., $T_L$ according to the UCI rules, or $A_L$  resulting from the sum of the   finishing times  of the 3 fastest riders (excluding their possible bonuses and penalties) of the team($\#$);  the last two columns report the team hierarchy according to the ascending value of  $\Delta_L^{(\#)}\equiv A_L ^{(\#)}- T_L^{(\#)}$.
 } \label{Table3TLALTdF23teams} 
 \end{center}
 \end{table}

   \begin{table}\begin{center}
\begin{tabular}{|c||c|c|c||c|c|c|c||c|c|c|c|} 
 \hline 
&\multicolumn{6}{|c|}{$VaE$ 2023}   \\	\hline
 \hline
 rank &  team 		& $T_L^{(\#)}$ &  team 		&  $ A_L^{(\#)} $  &team& $\Delta_L^{(\#)}$    \\ \hline \hline      
1	&	UAD	&	56:39:07	&	FST	&	57:27:16	&	TJV	&	0:36:06	\\
2	&	FST	&	56:45:07	&	UAD	&	57:29:18	&	SDW	&	0:36:39	\\
3	&	CSR	&	56:46:21	&	SDW	&	57:33:51	&	TFS	&	0:37:11	\\
4	&	SDW	&	56:57:12	&	CSR	&	57:33:55	&	AUB	&	0:38:04	\\
5	&	MOV	&	57:04:05	&	TJV	&	57:45:23	&	TIB	&	0:38:21	\\
6	&	TJV	&	57:09:17	&	MOV	&	57:46:37	&	FST	&	0:42:09	\\
7	&	DSM	&	57:10:52	&	TFS	&	57:53:19	&	COG	&	0:42:26	\\
8	&	TFS	&	57:16:08	&	DSM	&	57:53:29	&	LIV	&	0:42:31	\\
9	&	JAY	&	57:32:00	&	JAY	&	58:15:12	&	MOV	&	0:42:32	\\
10	&	COG	&	57:36:22	&	COG	&	58:18:48	&	DSM	&	0:42:37	\\
11	&	LIV	&	57:43:19	&	LIV	&	58:25:50	&	JAY	&	0:43:12	\\
12	&	LKF	&	57:50:32	&	TIB	&	58:33:59	&	EIC	&	0:46:11	\\
13	&	TIB	&	57:55:38	&	AUB	&	58:53:35	&	CSR	&	0:47:34	\\
14	&	EIC	&	58:07:34	&	EIC	&	58:53:45	&	UAD	&	0:50:11	\\
15	&	AUB	&	58:15:31	&	LKF	&	59:06:29	&	CDR	&	0:51:51	\\
16	&	BDU	&	58:31:27	&	BDU	&	59:40:22	&	FBW	&	0:52:25	\\
17	&	MAT	&	59:05:37	&	MAT	&	60:00:16	&	MAT	&	0:54:39	\\
18	&	BPK	&	59:15:03	&	BPK	&	60:17:18	&	STC	&	1:01:27	\\
19	&	HPU	&	59:24:51	&	HPU	&	60:42:30	&	BPK	&	1:02:15	\\
20	&	FBW	&	60:10:45	&	FBW	&	61:03:10	&	BDU	&	1:08:55	\\
21	&	CDR	&	60:27:06	&	CDR	&	61:18:57	&	LKF	&	1:15:57	\\
22	&	STC	&	61:17:50	&	STC	&	62:19:17	&	HPU	&	1:17:39	\\
23	&	SWT	&	...		&	SWT	&	...		&	SWT	&		...	\\
     \hline
\end{tabular}
\caption{Table displaying the  final ranking of  the ($M_L=22$) teams after the last $VaE$ 2023  stage, i.e., $T_L$ according to the UCI rules, or $A_L$  resulting from the sum of the   finishing times  of the 3 fastest riders (excluding their possible bonuses and penalties) of the team ($\#$);  the last two columns report the team hierarchy according to the ascending value of  $\Delta_L^{(\#)}\equiv A_L ^{(\#)}- T_L^{(\#)}$. SWT finished the $VaE$ 2023 multi-stage race with only 2 riders.
 }\label{Table5TLALVaE23teams} 
 \end{center} 
 \end{table}

   Together with Table \ref{TableTdFGdVaEILn0M0_MandF}, Table \ref{TableTLALDL23Fteamtimescharacteristics} {\it a posteriori} allows to compare race difficulties. It is easily observed that the time distributions of $T_L$ and $A_L$ are similar for $GdI$ and $TdF$, both races taking much more time than $VaE$, since indeed they are $\simeq 1.3$ longer. In the 3 cases, the skewness is positive $\simeq 0.8$, indicating a long tail in the final time distributions for the slowest teams.  The negative kurtosis for $GdI$ and $VaE$ indicate a flatter distribution, whence a race where teams find a more balanced competition, in contrast to the $TdF$ which presents a peaked distribution at the mean, -  itself close to the median. The same deduction holds when observing the $\sigma$, much shorter in $TdF$, indicating a fiercer competition between the top teams.

   \begin{table}
   \fontsize{9pt}{9pt}\selectfont
\begin{tabular}{|c||c|c|c||c|c|c||c|c|c|} 
 \hline 
&\multicolumn{3}{|c||}{$GdI$} &\multicolumn{3}{|c||}{$TdF$}  &\multicolumn{3}{|c|}{$VaE$}   \\	\hline
	& $T_L^{(\#)}$ &  $ A_L^{(\#)} $ & $\Delta_L^{(\#)}$   & $T_L^{(\#)}$ &  $ A_L^{(\#)} $ & $\Delta_L^{(\#)}$  & $T_L^{(\#)}$ &  $ A_L^{(\#)} $ & $\Delta_L^{(\#)}$  \\ \hline \hline   
$M_L$&\multicolumn{3}{|c||}{$23$} &\multicolumn{3}{|c||}{$22$}  &\multicolumn{3}{|c|}{$22$}   \\	\hline
Min.&	73:54:51	&	73:59:03	&	0:00:33	&	76:17:38	&	76:26:34	&	0:02:12	&	56:39:07	&	57:27:16	&	0:36:06	\\	
Max.	&	79:09:11	&	79:30:07	&	1:07:18	&	79:25:34	&	80:04:13	&	0:52:33	&	61:17:50	&	62:19:17	&	1:17:39	\\	\hline
$\mu$	&	75:39:04	&	75:54:54	&	0:15:50	&	77:26:01	&	77:44:41	&	0:18:39	&	58:08:16	&	58:57:51	&	0:49:35	\\	
$\gamma$&	75:38:07	&	75:53:53	&	0:10:37	&	77:25:50	&	77:44:25	&	0:12:58	&	58:07:26	&	58:56:53	&	0:48:15	\\	
Med.	&	75:08:40	&	75:17:26	&	0:12:33	&	77:27:29	&	77:46:23	&	0:15:02	&	57:46:56	&	58:29:55	&	0:44:42	\\	
$\sigma$	&	1:35:42	&	1:39:20	&	0:14:27	&	0:43:14	&	0:50:44	&	0:16:04	&	1:18:06	&	1:25:21	&	0:12:29	\\	
\hline
Skewn.	&	0.84232	&	0.88415	&	2.03400	&	0.70337	&	0.76641	&	1.06135	&	0.93943	&	0.84955	&	1.00163	\\
Kurt.	&	-0.59051	&	-0.38866	&	5.00210	&	0.77460	&	0.77392	&	-0.16262	&	-0.06882	&	-0.37660	&	-0.07262\\
\hline
 \end{tabular}
\caption{Table displaying  a few statistical characteristics of the final times (as defined in the text) distributions of teams in the 3 races: all times in h:m:s; $\mu $ is the arithmetic mean; $\gamma$ is the geometric mean; $\sigma$ is the standard deviation; Med. is the median; Skewn. the skewness; Kurt. the kurtosis.
} \label{TableTLALDL23Fteamtimescharacteristics}.   
 \end{table}

\subsection{Leadership Gap index}\label{leadershipgap}

Since  the measures  $ A_L ^{(\#)}$ and $ T_L^{(\#)}$ indicate a different team hierarchy, one can consider that their relative value: 
\begin{equation}\label{DeltaLeq}
 \Delta_L^{(\#)}\equiv A_L ^{(\#)}- T_L^{(\#)}\;,
 \end{equation}
which measures a behavioral difference between teams and/or riders performance 
 which might be due to team members skills or to coaching strategy.
 
 The  values of  $ \Delta_L^{(\#)}$ are given in Tables \ref{Table4TLALGdI23teams} - \ref{Table5TLALVaE23teams}. The smallest  $\Delta_L^{(\#)}$ 
  should correspond to the cases in which the  3 fastest riders in each stage remain  so throughout the whole competition, and finish it.  A large value in contrast indicates a team emphasis on a distributive role according to riders skills.  
 
 Indeed, the indicator reaches a large value if the riders  
  are not much concerned by their final rank. In this case, the “team leaders" do not seem to be “pre-defined". Thus, $ \Delta_L^{(\#)}$ appears to be a measure of a specific rider leadership definition in a team; in other words, the value of $ \Delta_L^{(\#)}$   
  measures a “gap" between team strategies, - depending on riders skills and coaches mandates.

\subsection{Stage and Race Temperature index}\label{ShannonStageTemperature}

Shannon and  Boltzmann-Gibbs entropy are analogous measures of disorder in informatics and thermodynamics.$^{\cite{Shannon,refTsallis}}$ The maximum entropy value corresponds to a state of maximum uncertainty, i.e., when all outcomes are equally likely,  pointing to a lack of structure or even predictability because of the absence of disorder. Thus, the concept seems of interest for measuring some operational effect in sport results.$^{\cite{Silvaetal2016}}$

Let $p_k ^{(s)}\;$ be the  relative time measure ($\simeq$ “contribution")  of the (3 fastest riders of a) team ($k$) to the total (cumulative)  time that was needed by the 3 fastest riders of each team  among all ($M_0$) teams in competition in order to finish the stage $s$

 
\begin{equation}\label{p_k}
p_k ^{(s)}\;= \frac{t_{k}^{(s)} }{\sum_{k=1}^{M_0}t_{k}^{(s)}} \;;
\end{equation}
$t_{k}^{(s)} $ is defined in terms of $ t^{(\#)}_{i,s}$, the finishing time of  one of the 3 fastest riders ($i=1,2,3$)  of  team $k$ (or $\#$, in terms of UCI codes)  for that stage; see Eq. (\ref{tsteameq}).
 This measure can be taken per definition like a probability of a team best finishing time among others. Thus,  $p_k ^{(s)}$ can serve as a characteristics how rare the occurrence of such  an outcome is.
  The stochastic Shannon entropy reads
\begin{equation}\label{p_i}
S_k ^{(s)}= -\; p_k ^{(s)}\; ln (p_k ^{(s)}) \;.
\end{equation}
 
The average of such a $S_k ^{(s)}$ over the total probability distribution leads to the Shannon information  entropy;
\begin{equation}\label{entropyS(s)}
S_{s}^{(\#)}\;\equiv\; <S_k ^{(s)}>\;\equiv \;-\; \sum_{k=1}^{M_0}\; \;{p_k ^{(s)}\;ln(p_k ^{(s)})}  \;.
\end{equation}

The  whole race entropy $S_L$   derives from summing over all the teams entropy: $S_L=\sum_{\#} S_s ^{(\#)}$.

Thereafter, reconnecting the Shannon information entropy to the thermodynamic Boltzmann-Gibbs entropy, one can define a team dependent  (“generalized") temperature during the stage $s$ as
\begin{equation}\label{stochastictemperature}
\theta_k ^{(s)}\;\simeq\;\frac{-1}{p_k ^{(s)}\;ln(p_k ^{(s)})} \;.
\end{equation}
{\it Mutatis mutandis}, this is analogous to the “temperature of financial markets".$^{\cite{refKozuki03,refLietal23}}$ 
 In fact, one may propose a ($s$-th) “stage temperature index" as
\begin{equation}\label{temperature}
\theta^{(s)} \; \simeq\;  \frac{-1}{\sum_{k=1}^{M_0}\;{p_k^{(s)}\;ln(p_k ^{(s)})}} \;\;\; ;
\end{equation}
  the smaller  it is, the cooler appears to be the  competition during the stage. Indeed, recall that the Shannon  entropy of a uniform distribution, i.e, if all  $ p_k$ are equal, thus in the absence of disorder,  is   the maximum entropy value which can occur. Randomness or disorder, in  $ p_k$'s,  thereby corresponds to a high “(behavioral) temperature", here seen as an intense competition.  

 The “team temperature" is forecasted to be higher if the riders have much  strategic freedom. It is readily expected that such a temperature is lower if leaders are well defined. In the present case, this occurs, as easily understood,  if $\Delta_L$ is small. 
 
 The overall race temperature is of course
\begin{equation}\label{temperature_L}
\theta_L\; = \frac{1}{S_L} =    \frac{-1}{\sum_{s=1}^L \; \sum_{k=1}^{M_0}\;{p_k^{(s)}\;ln(p_k ^{(s)})}} \;\;\; .
\end{equation}
where 
the latter $\theta_L$ has been calculated from the time of riders irrespective of the fact that they might not have finished the whole race; thus, more exactly, one should have written the $\theta_L$ rather as $\theta_L(T_L)$.  Recall that instead of $t_{k}^{(s)}$ in Eq.(\ref{p_k}), one can  consider that the pertinent time is that of   riders finishing  the whole race. Following the above path, one would obtain a final race temperature $\theta_L(A_L)$. Moreover, another temperature can be derived from $A_L-T_L$ data, leading to some $ \theta_L(T_L-A_L)$. Due to the nonlinear  data transformations,   $ \theta_L(T_L-A_L)$  $\neq$  $ \theta_L(T_L)-  \theta_L(A_L)$, - in contrast to the Entropy which has an additivity property.$^{\cite{refTsallis}}$   
 
Again, one can justify the semantic validity for calling  $\theta_L$ a relative temperature index. Indeed,  $\theta_L$  appears to be a measure of the distribution of riders kinetic energy at the end of a race (or stage).  
   
\section{Other  Indicators} \label{Others}

\subsection{Atkinson index }\label{AC}
 
Thus, the Atkinson index $At_L$ can be used for evaluating the strength dispersion of teams in a race.$^{\cite{Atkinson1970}}$
Per definition,
 \begin{equation}\label{AtI}
At_L = 1 - \gamma/\mu \;,
\end{equation}
where $\gamma$ is the geometric mean and $\mu$ is the arithmetic mean of the distribution, - reported in Table \ref{TableTLALDL23Fteamtimescharacteristics}.  The Atkinson index has previously been used in sport in order to measuring competitive balance with an application to English Premier League football.$^{\cite{BorooahManganfootball}}$  One can obviously apply the notion to the present cases. It could also be considered for daily stages ($s$) rather than the whole race ($L$), but such an application is left for further work.

\subsection{Coefficient of Variation }\label{CV}
 Among the indicators using statistical characteristics of distributions, the Coefficient of Variation ($CV$)  measures the data relative dispersion,  i.e., pointing to the dispersion ($\sigma$) around the mean ($\mu$) of the distribution; thus, expressed in percentage it is somewhat hinting to inequalities.  Per definition, one has
 \begin{equation}\label{CVeq}
CV_L=  \sigma / \mu,
\end{equation}
 easily obtained from the distributions characteristics.

 \subsection{Herfindahl-Hirschman index} \label{Herfindahl}

 One may also recall the Herfindahl-Hirschman index serving to measure the “amount of competition" between economic entities,$^{\cite{Brezinaetal2016,YigitTur2012,OladimejiUdosen2019,Handoyoetal2023}}$
 - or for our examples, between  teams.$^{\cite{ausloos2020concentrationTdF}}$

The Herfindahl index, also known as Herfindahl-Hirschman index ($HHI$), is  a measure of “concentration in a market".$^{\cite{Hirschman}}$  Formally, in obvious notations, it reads
\begin{equation}\label{HHIeq}
HHI =  \sum_{i=1}^{N} \left(\frac{y_k}{\sum_j y_j} \right)^2,
\end{equation} 
where $y_k$ is some economic measure, like a company size, or its share (thus, a concentration) in a market.
Thus, a $HHI$ index $\leq 0.01$  indicates a highly competitive market between $N$ firms.
From a portfolio point of view, a low $HHI$ index implies a very diversified portfolio; a high concentration demands $HHI \ge 0.25$; a low concentration $HHI \le 0.15$; $HHI$ ranges between $1/N$ and 1.$^{\cite{Brezinaetal2016}}$

Adapted  to the case of sport team ranking, $y_k$ can be considered as the finishing time  ($t_k$) of a team ($k$), i.e.,  leading to $HHI = \sum HHI_k \equiv \sum p_k^2$, as defined in Eq. (\ref{p_k}), and where the relevant team times are selected depending on the chosen   $ A_L ^{(\#)}$ or $ T_L^{(\#)}$ scheme, or even $\Delta_L$. 

As an extreme example, - which sometimes occurs, if 3 riders of each team arrive together, whence have the same finishing time for a stage, all terms in the Eq.(\ref{HHIeq}) sum are equal, whence $HHI=1$, pointing to uniformity or in other words to a rather weakly competitive race. In other words, an increase in $HHI$ represents a decrease in competitive balance.$^{\cite{OwenRyanW07,OwenOwen22}}$

 One  sometimes says that the “number of effectively important competitors" is the inverse of the Herfindahl index.

 A normalized $HHI$  is sometimes used in order to attempt some universal definition:   \begin{equation}\label{NormHH}
HHI^{*} =  \frac{\left (HHI - \frac{1}{N} \right )}{ 1-\frac{1}{N} }.
\end{equation}
with the appropriate $N$; it ranges between 0 and 1.

\subsection{Gini Coefficient } \label{Gini}
 
 The most popular way for quantifying  inequality levels,   in  socio-economic systems, is through the Gini coefficient ($GiC$).$^{\cite{MarmaniKaur,QQ49.15.2307RCMAstatistical}}$
It reads  \begin{equation}\label{Ginieq}
GiC =   \frac{1}{<y>} \frac{\sum_{i=1}^N \; \sum_{j=1}^N |y_i\;-\;y_j |}{2\;N^2} 
\end{equation}
where  the $i$-th item has a measure $y_i$, and $<y>$ is the average value of this quantity over the whole set of $N$ elements. In the present case $y_i$ can be the resulting  time ($t_k$) of a team ($k$) due to 3 riders as above within the $T_L$ or $A_L$ schemes. 
 The Gini coefficient should be equal to 0 if  all teams   are equivalent but, e.g., = 1 if one team is much above others, or in socio-economic terms, which would be “monopolizing the whole of the available resources".  One should expect a $GiC \simeq 0$ if the competition  has no winner, - in other words if teams have equivalent final “values".

 \subsection{Theil index} \label{Theil_L}
  
For completeness, one can define the “final" Theil  index. One has
\begin{equation} \label{eq:IdA}
 Th_L = \frac{1}{M_L}\; \sum_{k=1}^{M_L}  \frac{x_k}{\langle x_k\rangle} \ln \frac{x_k}{\langle x_k \rangle},
\end{equation}
summing $\frac{x_k}{\langle x_k \rangle}$ over the different  (finishing) teams $k$ in the  race and where  $\langle x_k\rangle $ is the  mean value, of any variable, which here can be any $t_k$.  
  This transformation  induces negative and positive values of the (log-transformed) data, depending on the ratio $x_k/<x_k>$$\equiv$ $Th_k$. Whence, $Th_L\equiv \sum Th_k$ can be very small.

\subsection{Pietra-Hoover index }\label{PRS}

 It seems of interest, for emphasizing the structure, like the maximum position and the corresponding percentage of the relevant population, to display the data as the difference between the Lorenz curve ($LoC$) and the line of perfect equality.$^{\cite{Lorenz1905,FENSRszeszowAPPA129}}$  One has 
\begin{equation}\label{Ginidistance}
\delta h_{k} = \;\frac{1}{N}\; [\sum_{j =1}^k\;j\;y_j - k ]
\end{equation}
with $ k \le N$.
 In fact, this  is the Pietra-Hoover (inequality) index.$^{\cite{JosaAguado2020,Frosini2012}}$
   \begin{equation}\label{PHIeq}
 PHI= \frac{\sum_i^N\; g_i-<g_i>} {2 \sum_i^N\;g_i}\;.
    \end{equation}
It indicates how the variable values should be (re)distributed in order for them to create a perfect equality in times.     High values of the index obviously represent a high inequality level since a greater redistribution of values is required in order to achieve equality; vice-versa, lower values of the index represent a lower inequality level.  

\subsection{Rosenbluth  Coefficient }\label{Rosenbluth}
 The Rosenbluth  Coefficient 
   is defined as 
 \begin{equation}\label{Rosenbluthindex}
     RoC=\frac {1}{2\sum is_{i}-1}  
        \end{equation} where the symbol $i$ usually indicates a firm's rank position on economic markets.$^{\cite{HallTideman67}}$
          Thereafter, $s_i$  can be taken as the rank of the percentage of a size measure,  
like  some $t_k$  ratio. 
    
    Practically, the Rosenbluth index assigns more weight to weaker competitors.      Such a  measure which weights each competitor by its rank rather than by its “share" seems very  appealing for our purpose.

The Rosenbluth coefficient is related to the Gini coefficient through
   \begin{equation} \label{RIeq}
RoC= \frac{1}{N\;(1-GiC)} \; .  
     \end{equation}

\begin{figure}[ht] 
\includegraphics[height=14cm,width=14cm]{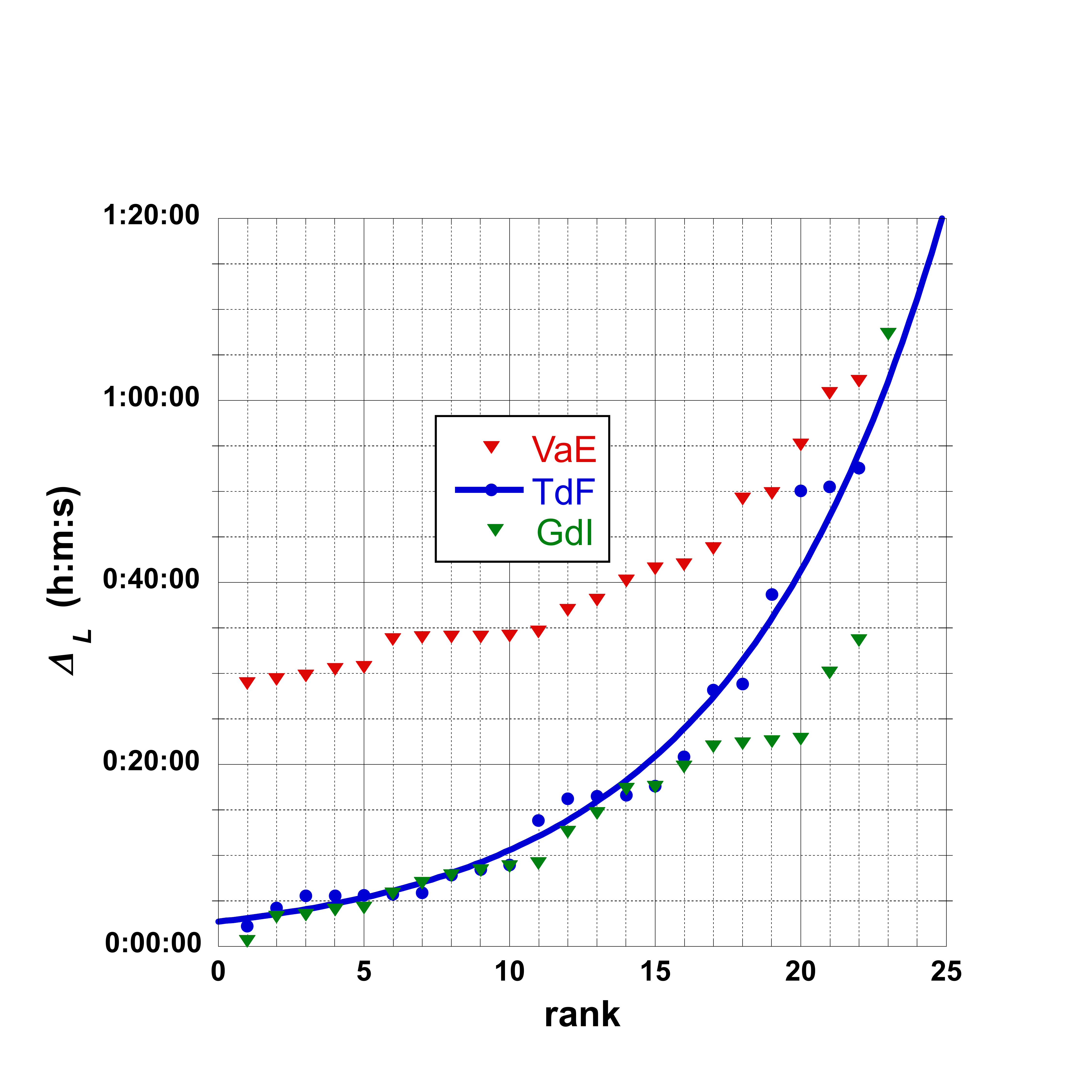}
\caption{
Plot of      $\Delta_L^{(\#)}\equiv A_L ^{(\#)}- T_L^{(\#)}$(h:m:s), measuring the difference between the suggestion$^{\cite{EJOR}}$   demanding  team rankings when only considering riders who  finish the race in and the UCI rules  for teams final time  ranked in time increasing order, according  to UCI ($T_L$)  rules;
 colors correspond to Giro d'Italia $GdI$ (green), Tour de France $TdF$ (blue),  Vuelta a España   $VaE$ (red), -  2023 female races.   For $TdF$, the (best OLS) fit leads to :
$y$ = 162.49$\;$$e^{(0.1362\;x)}$;   $R^2$= 0.9752.}
\label{Plot 3 D-L T-L d-h-m} 
\end{figure} 
 
  \begin{table}
   \fontsize{9pt}{9pt}\selectfont
   \begin{tabular}{|c||c|c|c||c|c|c||c|c|c|}  
 \hline 
&\multicolumn{3}{|c||}{$GdI$} &\multicolumn{3}{|c||}{$TdF$}  &\multicolumn{3}{|c|}{$VaE$}   \\	\hline
	& $T_L^{(\#)}$ &  $ A_L^{(\#)} $ & $\Delta_L^{(\#)}$   & $T_L^{(\#)}$ &  $ A_L^{(\#)} $ & $\Delta_L^{(\#)}$  & $T_L^{(\#)}$ &  $ A_L^{(\#)} $ & $\Delta_L^{(\#)}$  \\ \hline \hline   
$M_L$&\multicolumn{3}{|c||}{$23$} &\multicolumn{3}{|c||}{$22$}  &\multicolumn{3}{|c|}{$22$}   \\	\hline
$\theta_L$ &	0.31895	&	0.31895	&	0.35652	&	0.32352	&	0.32352	&	0.36126	&	0.32354	&	0.32354	&	0.32652	\\\hline
$S_L$	&	3.13528	&	3.13527	&	2.80487	&	3.09100	&	3.09099	&	2.76806	&	3.09080	&	3.09077	&	3.06258	\\
$At_L$	&	1.05e-04	&	1.13e-04	&	0.16414	&	2.06e-05	&	2.81e-05	&	0.15940	&	1.18e-04	&	1.38e-04	&	0.01386\\
$CV_L$	&	0.02062	&	0.02133	&	0.89212	&	0.00909	&	0.01062	&	0.84168	&	0.02187	&	0.02357	&	0.24586	\\ \hline
$HHI$	&	0.04350	&	0.04350	&	0.07808	&	0.04546	&	0.04546	&	0.07766	&	0.04548	&	0.04548	&	0.04820	\\
$HHI^*$	&	1.93e-05	&	2.07e-05	&	0.03618	&	3.94e-06	&	5.38e-06	&	0.03373	&	2.28e-05	&	2.65e-05	&	0.00288	\\
$GiC$	&	0.01120	&	0.01160	&	0.43649	&	0.00500	&	0.00582	&	0.44368	&	0.01192	&	0.01291	&	0.13211	\\
$Th_L$	&	2.11e-04	&	2.26e-04	&	0.33062	&	4.12e-05	&	5.63e-05	&	0.32298	&	2.38e-04	&	2.76e-04	&	0.02846	\\
$PHI$	&	0.00867	&	0.00883	&	0.32024	&	0.00359	&	0.00422	&	0.33864	&	0.00889	&	0.00984	&	0.09997	\\
$RoC$	&	0.04397	&	0.04399	&	0.07716	&	0.04568	&	0.04572	&	0.08171	&	0.04600	&	0.04605	&	0.05237	\\
\hline
 \end{tabular}
\caption{Table displaying  a few concentration coefficients resulting from the  final ranking of  the  ($M_L$) teams at the end of the $GdI$, $TdF$, and $VaE$ 2023  race, for $T_L$,  $A_L$,   and $\Delta_L^{(\#)}\equiv A_L ^{(\#)}- T_L^{(\#)}$.  $M_L$ is  the Number of finishing teams;  $S_L$ is the  final Entropy;  $At_L$	 the Atkinson index; $CV_L$  the Coefficient of Variation; $HHI$  the Herfindahl	  index;   $HHI^*$   the Normalized Herfindahl  index; $GiC$  the Gini coefficient;  $Th_L$  the final Theil  Index; $PHI$  the Pietra-Hoover	coefficient; $RoC$  the Rosenbluth coefficient. One distinguishes global measures (top of Table) from those depending on the ranking order  (bottom of Table).} \label{TableTLALGdI23Fteamsconcentrations}.   
 \end{table}

\section{Results and Analysis}\label{ResultsandAnalysis}
 
 Numerical results  should be examined along two perspectives: (i) one takes into account new indices based on imposing the constraint that a team evaluation  and ranking depends on the members at the valuation time (here at the end of the race),  but
 (ii)  besides global statistical values, i.e., irrespective of the team rank, one distinguishes values taking into account team ranks, as a weight. 
The  global values are found in Table  \ref{TableTLALGdI23Fteamsconcentrations},  in the top and bottom respectively.
Most of the outputs arise from freely accessing $https://www.wessa.net/desc.wasp$.

One can remark that the orders of magnitude  for $T_L$ and $A_L$ do not differ much, but these differ from $\Delta_L$.  The entropy and the leadership gap temperature significantly differ from race to race. This can be tracked to the similar order of magnitude of the $p_k$, implying similarities in $\theta_L$. This indicates that the overall distribution of rankings has not much influence on  global  characteristics; in fact, one should expect that the $teams$ hierarchies do not much differ from race to race.    
Concerning the rank effect as a weight for calculating indicators, one observes the largest effects in the “unweighted" indicators. This suggests to provide displays based on ranks.

The first display of interest should be the  new $\Delta_L$ indicator variation as a function of the rank for the different races.  Fig. \ref{Plot 3 D-L T-L d-h-m}  shows a plot of $\Delta_L$ team final times  ranked in time increasing order, according  to UCI  rules. Once and for all notice the meaning of colors: they correspond to the Giro d'Italia ($GdI$, green triangles), Tour de France ($TdF$, blue circles), and Vuelta a España   ($VaE$, red triangles).   For $TdF$, the (best OLS) fit leads to some smooth exponential behavior.  In  contrast, a  simple curve cannot fit the $GdI$ and $VaE$ data in which steps appear, suggesting team clustering, as discussed below. 

For the best ranked teams  ($r \le 16$),  $\Delta_L$ values for $GdI$ and $TdF$ are very similar. However the $VaE$ $\Delta_L$ values  are very different from those in the other two races, pointing to either different difficulties or/and   to different types of skills of teams members, and  subsequently strategies, as hinted in the previous Sections.  

\begin{figure}[ht] 
\includegraphics[height=14cm,width=14cm]{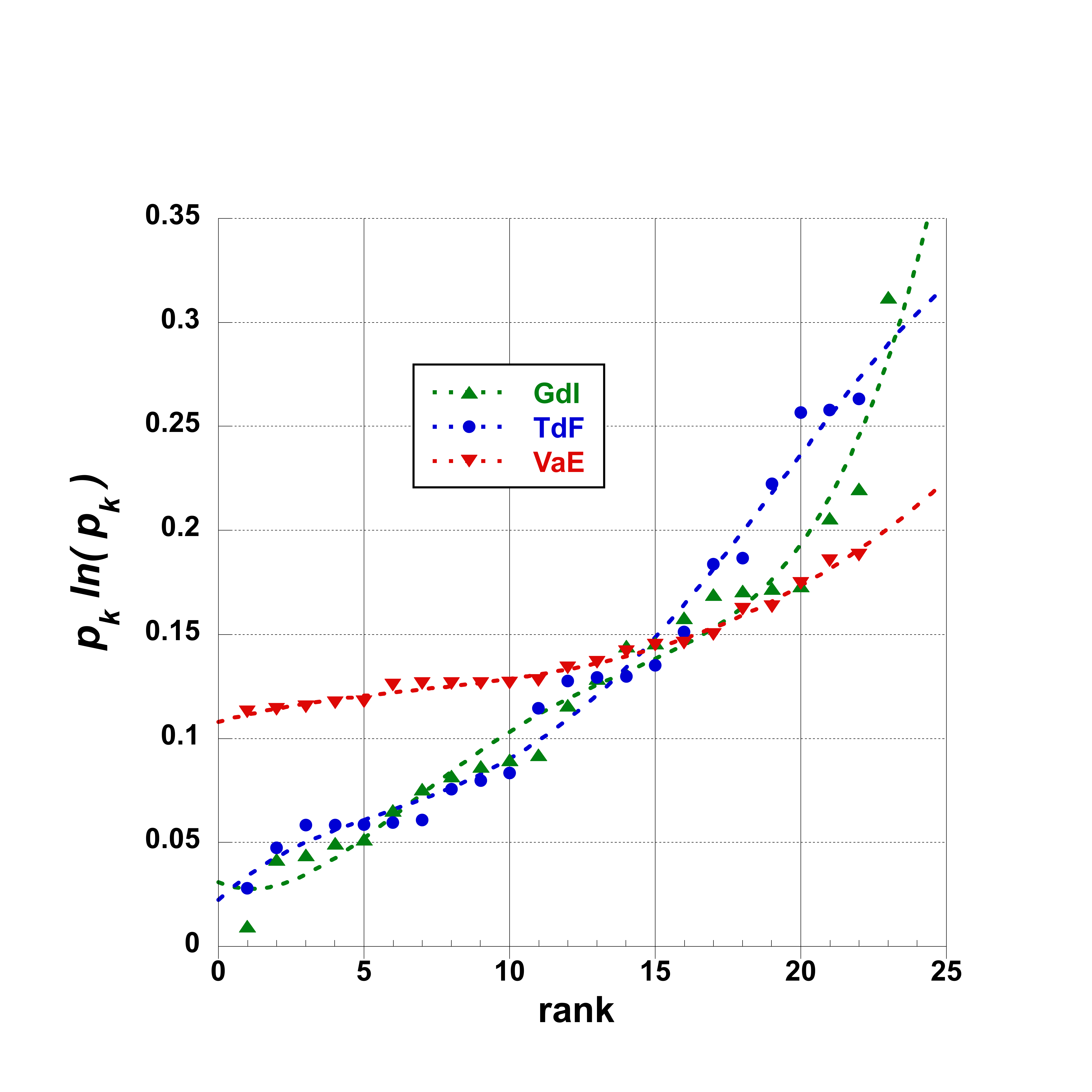}
\caption{
Plot of      the stochastic Shannon entropy of a team,  $p_k\;ln(p_k)$, - see Eq. (5),   for the   $\Delta_L^{(\#)}\equiv A_L ^{(\#)}- T_L^{(\#)}$(h:m:s), distribution, in ranked time increasing order;
 colors correspond to Giro d'Italia $GdI$ (green), Tour de France $TdF$ (blue),  Vuelta a España   $VaE$ (red) 2023 female races.   The dashed line is a 4-th order polynomial, used as a guide to the eye only, - for better distinguishing the rank evolutions  of    $p_k\;ln(p_k)$   index in the 3 races. } .
\label{Plot 13 plnp pol4 fit} 
\end{figure}

Other new indicators imply the $p_k$ rank distribution. Fig. \ref{Plot 13 plnp pol4 fit}  
 presents the plot of   $p_k\;ln(p_k)$, a term appearing in calculating $S_L$,    for the $\Delta_L$ distribution, in ranked time increasing order. A marked difference is found between data for $VaE$ and $GdI$ or $TdF$. The evolution is nevertheless rather similar, presenting 3 inflexion points.
 The dashed line is a 4-th order polynomial, used as a guide to the eye only,  for distinguishing the rank dependence of  the $p_k\;ln(p_k)$   indicator in the 3 races.

\begin{figure}[ht] 
\includegraphics[height=14cm,width=14cm]{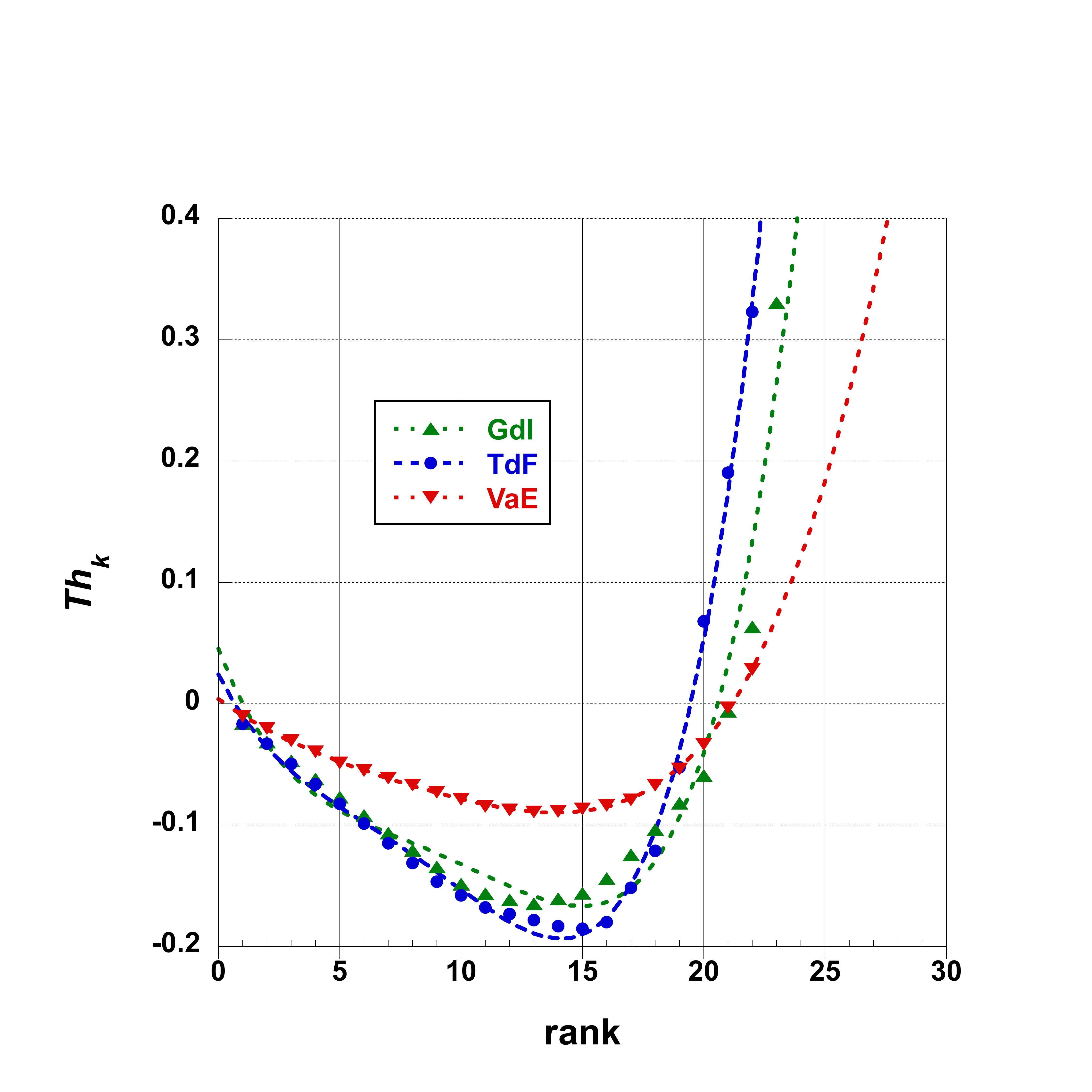}
\caption{
Plot of  the     team Theil index contribution, $Th_k$,    to the final Theil index, - see Eq. (16),  for the $\Delta_L$ distribution, in ranked time increasing order; 
 colors correspond to Giro d'Italia $GdI$ (green), Tour de France $TdF$ (blue),  Vuelta a España   $VaE$ (red) 2023 female races.   The dashed line is a 4-th order polynomial, as a guide to the eye only, and for better pointing the crucial (core) rank ($r_c$) at the minimum of  the $Th_k$  indicator. } .
\label{Plot 10 Th ranked pol4} 
\end{figure}

Similarly, one can study the contribution of each team to the Theil index through   $Th_k$ $\equiv$  $x_k/<x_k>$, 
 
Fig. \ref{Plot 10 Th ranked pol4} 
shows the plot of  the $Th_k$  values for the $\Delta_L$ distribution, in ranked time increasing order.   The dashed line is a 4-th order polynomial, as a guide to the eye only, pointing the crucial (core) rank ($r_c$) at the minimum of  the $Th_k$s  distribution. Indeed, a marked “difference" occurs for teams below $r \simeq 15$, in the 3 races. A similar behavioral aspect is found for the variation of the $HHI_k$s, - not shown for saving space.
 
\begin{figure}[ht] 
\includegraphics[height=14cm,width=14cm]{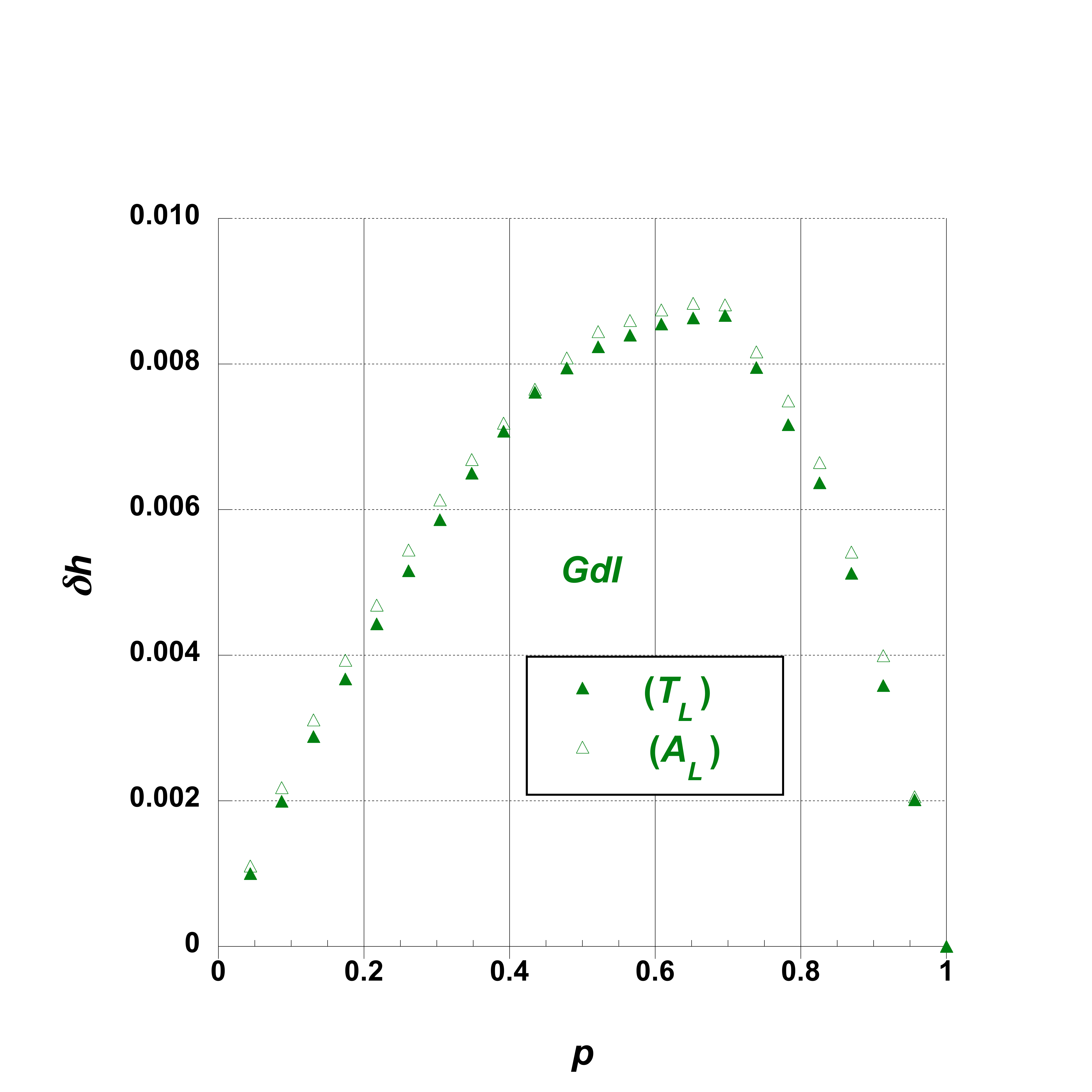}
\caption{
Display of the distance $\delta h$ between the Lorenz curve and the line of perfect equality, on a Gini coefficient graph, for the distribution of $T_L$    or $A_L$ times of  the 2023 $GdI$ finishing teams; the maximum  of  each curve gives the relevant crucial (core) rank; $p$=1 corresponds to $M_L$ = 23 for $GdI$.    Recall that for the $TdF$ and $VaE$, $M_L = 22$, whence see Fig. \ref{Plot 53 dh TL AL TdF VaE}.}
\label{Plot 52 dh TL AL GdI} 
\end{figure}

\begin{figure}[ht] 
\includegraphics[height=14cm,width=14cm]{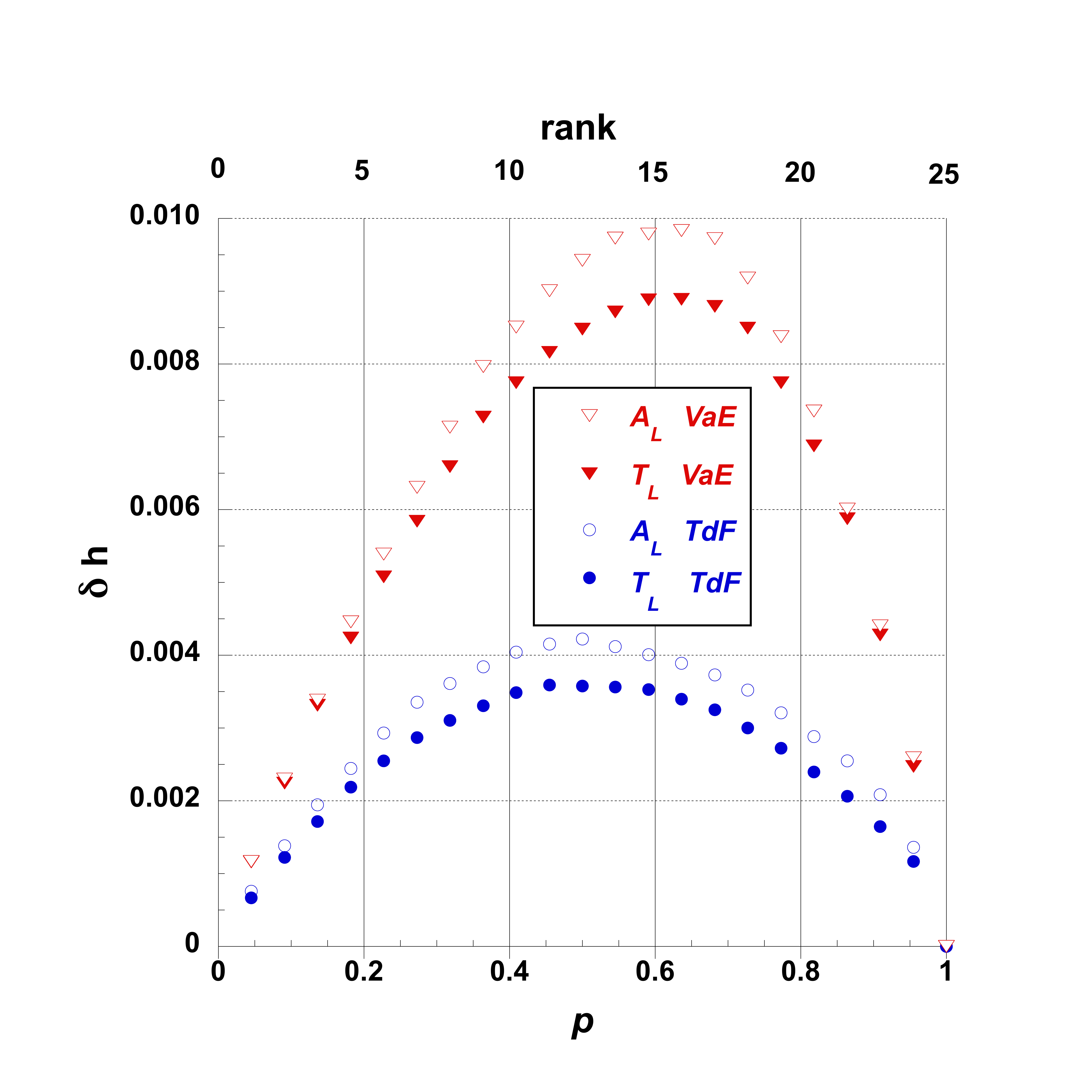}
\caption{
Display of the distance $\delta h$ between the Lorenz curve and the line of perfect equality, on a Gini coefficient graph,  for the distribution of $T_L$    or  $A_L$  times distributions; the maximum  of  each curve gives the crucial (core) rank; $p$=1 corresponds to $M_L$ = 22 for $TdF$ and $VaE$; symbols correspond to Tour de France $TdF$ (blue; circles) and Vuelta a España   $VaE$ (red; triangles) respectively, -  2023 female races.     Recall that for $GdI$, $M_L = 23$, whence see Fig. \ref{Plot 52 dh TL AL GdI}.}
\label{Plot 53 dh TL AL TdF VaE} 
\end{figure}

The most classical indicator of inequalities is the Gini coefficient; it is often presented as resulting from the ratio of surfaces, - somewhat difficult to estimate at first sight.   In order to provide a better vizualisation, one displays the evolution of the distance to the equality distribution line $\delta h$ as a function of the rank:
 Fig. \ref{Plot 52 dh TL AL GdI} 
  displays   $\delta h$   for the distribution of $T_L$ and $A_L$ times only for  the 2023 $GdI$ finishing teams;
  on 
Fig. \ref{Plot 53 dh TL AL TdF VaE}, 
one finds the distance $\delta h$ for the distribution of $T_L$ and $A_L$  times distributions for the 2023 $TdF$ and  $VaE$  female races, respectively.

\begin{figure}[ht] 
\includegraphics[height=14cm,width=14cm]{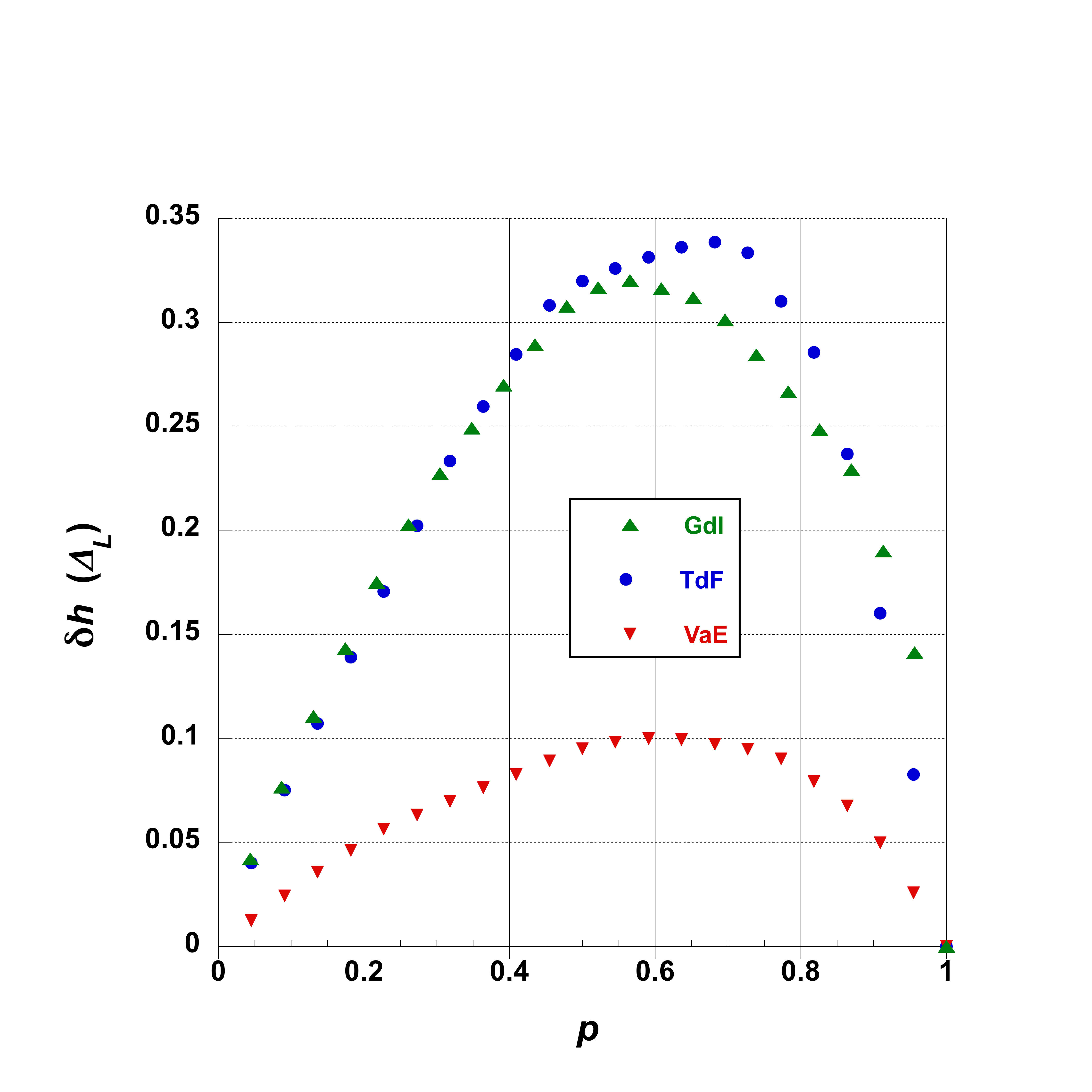}
\caption{
Display of the distance $\delta h$ between the Lorenz curve and the line of perfect equality, on a Gini coefficient graph,   strictly  for the $\Delta_L$ distribution of relative times; the maximum  of  each curve    indicates the  crucial (core) rank    between the top and bottom teams       thereby suggesting different team values; $p$=1 corresponds to $M_L$ = 23 for $GdI$ and = 22 for $TdF$ and $VaE$; colors correspond to Giro d'Italia $GdI$ (green), Tour de France $TdF$ (blue),  Vuelta a España   $VaE$ (red), -  2023 female races }
\label{Plot 51 dh DL GdI TdF VaE} 
\end{figure}

Fig. \ref{Plot 51 dh DL GdI TdF VaE} 
displays  the distance $\delta h$ between the Lorenz curve and the line of perfect equality, for the distribution of relative times.  The maximum  of  each curve gives the crucial (core) rank, since $p$=1 corresponds to the highest rank, $M_L$ = 23 for $GdI$ and = 22 for $TdF$ and $VaE$.  
   It is remarkable that  the maximum  of  each curve occurs near the crucial (core) rank $r_c$ in all cases. 

   One may propose some interpretation of a team core existence by analogy with an anharmonic oscillator. A few teams, the main ones,  have definite goals and aims, with {\it ad hoc team composition} but the other teams anticipate or respond to the strategy of the leading teams which have well defined and expectedly well performing leaders. Beside the rider skills differences, whence anticipating different levels of performance, one can also imagine that the not-too-best teams response introduces some non-linearity in the overall race dynamics description.

   In all these Figures, in particular in Figs. \ref{Plot 13 plnp pol4 fit} and  \ref{Plot 51 dh DL GdI TdF VaE}, it might be noticed that the $VaE$ data points present some  different behavior than those for $GdI$ and $TdF$. Such different behaviors can be on one hand traced back to quite different time distributions: observe the orders of magnitudes of $GiC$ and  $HHI^*$, and {\it a fortiori}      
   $Th_L$, $At_L$, and $CV_L$, even taking into account the length difference of the races. On the other hand, this behavior likely reflects the difference in team approaches for the races, i.e., the team compositions and the teams levels (see Table \ref{TableUCIacronMandF}), leading to
   time distributions differences. 

\begin{figure}[ht] 
\includegraphics[height=14cm,width=14cm]{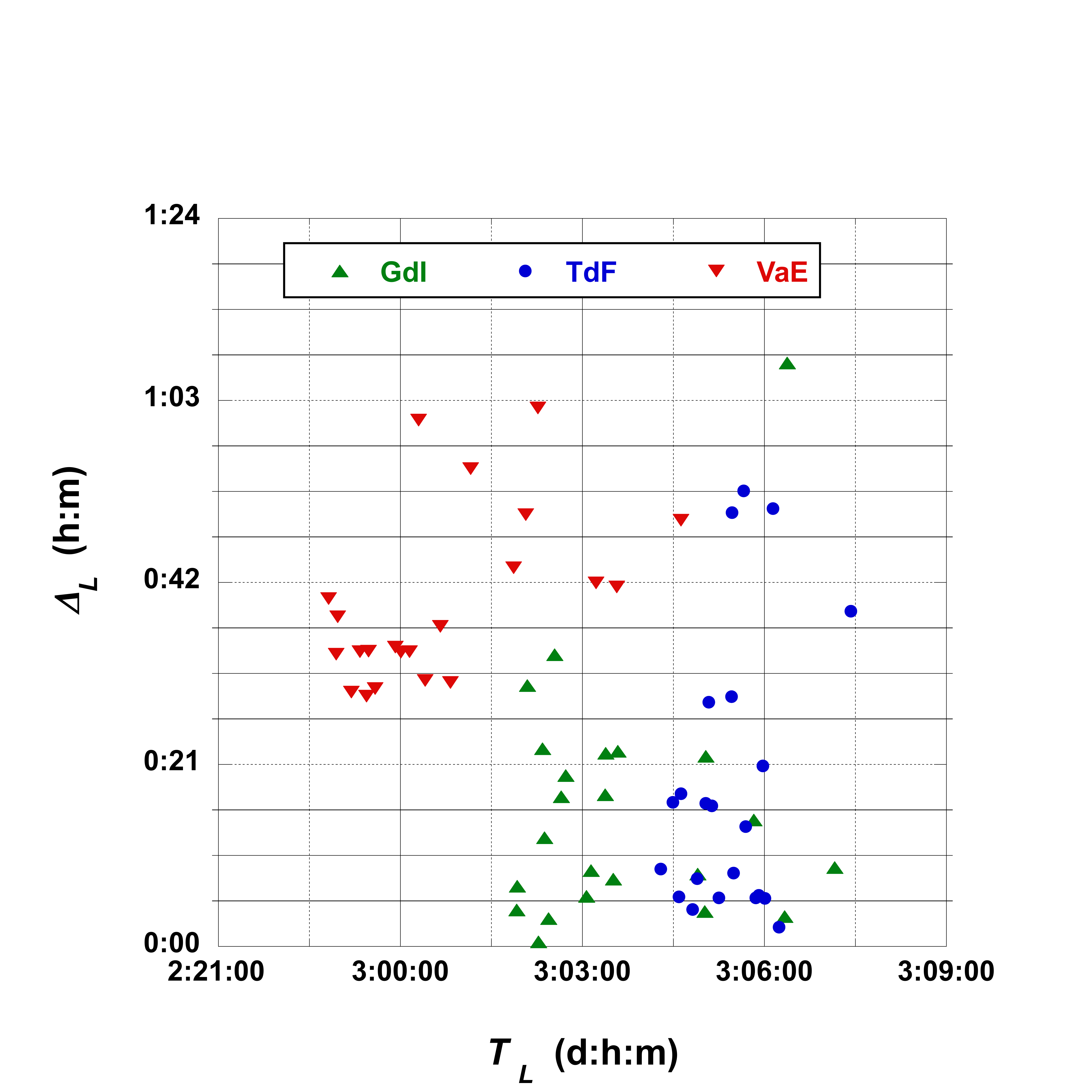}
\caption{
Scatter plot of        $\Delta_L^{(\#)}\equiv A_L ^{(\#)}- T_L^{(\#)}$, measuring the difference between  $A_L ^{(\#)}$, the suggestion$^{\cite{EJOR}}$   demanding  team rankings when only considering riders who  finish the race,  and $T_L ^{(\#)}$, according to the UCI rules  for teams final time   according  to UCI ($T_L$)  rules, vs. $T_L$  teams final time, ranked in time increasing order;  $\Delta_L$ and $T_L$  teams final time are given in   hours:minutes (h:m),   and days:hours:minutes (d:h:m),  respectively,  - the latter ranked in time increasing order, according  to UCI ($T_L$)  rule; colors correspond to Giro d'Italia $GdI$ (green), Tour de France $TdF$ (blue),  Vuelta a España   $VaE$ (red), - 2023 female races. The $VaE$ data has been rescaled, see text, to make the figure readable.   }
\label{Plot 2 D-L T-L d-h-m} 
\end{figure}

Fig. \ref{Plot 2 D-L T-L d-h-m} 
is a plot of $\Delta_L$ vs. $T_L$  teams final time, 
 according  to UCI   rule.  
  N.B. The $VaE$-$\Delta _L$ and $VaE$-$T_L$ data have been rescaled,  (divided and multiplied, respectively) by a factor 1.25, in order to display the data on the same  figure. Such a scaling  factor  ($\simeq 960/740$) roughly corresponds to the ratio between the lengths of the  relevant races; see Table \ref{TableTdFGdVaEILn0M0_MandF}.  A very similar     
 figure can be made for illustrating the $\Delta_L$ and $T_L$ relationship, without bringing anything specifically more interesting;  it is omitted for space saving.

   In the same line of thought, one could compare the differences in team ranks in $T_L$ and $A_L$ with respect to each other, besides discussing both vectors with respect to $\Delta_L$.  This can be made along the Kendall $\tau$ coefficient or measure their relative Kemeny distance. These considerations  are  left for Appendix A, in order to refer to the  “weighted preferences'' notions, somewhat of wider interest than the above measures.

Observing the Tables and the Figures, one is rightly tempted to look for team clustering, within the perspective of this study.  

(i) In  Tables \ref{Table4TLALGdI23teams}-\ref{Table5TLALVaE23teams}, one can observe clusters,  admitting rank swapping: 
  \begin{itemize}
\item $GdI$:  the best 2 teams are  swapped in $T_L$ and $A_L$;  but  the internal ranking is much scrambled between the 3-rd and 8-th team; the  following 15 teams are equally ranked, except the last 2;
 \item$ TdF$: scrambling of the best 2nd and 5-th teams; much scrambling thereafter, but with mere small swapping, in the center of the ranks,  up to the last ranks;
   \item $VaE$:  the first 8 teams swap ranks regularly by pairs; with almost no scrambling thereafter. 
\end{itemize}

(ii) In   Fig.  \ref{Plot 3 D-L T-L d-h-m} and   Fig. \ref{Plot 2 D-L T-L d-h-m}, one can observe clusters of teams for
 \begin{itemize}
 \item  $GdI$: a group of 16  below $T_L$ $\simeq$ 3:04:30 and a group of 8 above;  Fig. \ref{Plot 2 D-L T-L d-h-m};
\item  $GdI$: more precisely, below  $r_c = 11$  and above the rank for $r_e \ge 21$;
\item  but see also a step at $r_d \ge16$ separating 2 clusters;   Fig.  \ref{Plot 3 D-L T-L d-h-m};
\item  $TdF$:  below  $r_c \simeq 10$ and  others below and above $r_d=17$ ;   Fig.  \ref{Plot 3 D-L T-L d-h-m};
\item $TdF$:  a cluster of 16 teams centered on $T_L \simeq 3:02:00$;   Fig. \ref{Plot 2 D-L T-L d-h-m};
\item  $TdF$: a cluster of 4 teams appears above 00:31  from $\Delta_L$ values;  Fig. \ref{Plot 2 D-L T-L d-h-m};
 \item $VaE$:  5 clusters seem to appear  below   $r_c \le 5$,  $r_d \le 11$, and for $r_e \le 17$, on Fig.  \ref{Plot 3 D-L T-L d-h-m};
 \item $VaE$: a  cluster of 14 teams, around [$T_L; \Delta_L$] $\simeq$  [$2:23:00; 0:32$],   Fig. \ref{Plot 2 D-L T-L d-h-m}.
\end{itemize}

These observations remind of self-organized complex systems, often amounting to 3 clusters (high, medium, low ranks/classes) under collaboration-competition rules as found in many societies.$^{\cite{caram1,IJMPCFIFAMARCAGNV}}$

 \section{Discussion} \label{Discussion}
 
 Before discussing features, let it be recalled that one looks for (new) indicators containing (new) filters, in particular for stressing the contribution input of team members to a team rank, - due to manager selections and strategies. Thus, one introduces the “leadership gap"  $\Delta_L$, Eq. (\ref{DeltaLeq}),  and the “race temperature" $\theta_L$, Eq. (\ref{temperature_L}),  - beside  classical ranking indices. From the numerical values of interest in a set of study cases, one expects to deduce qualitative aspects of  wider insights. Indeed, the classically used indicators (Section \ref{AC} - \ref{Rosenbluth})  provide hypotheses (or assumptions) to managers devising strategies. The new indicators (Section \ref{leadershipgap} - \ref{ShannonStageTemperature}) increase perspectives. 
 
From all Figures and Tables, 
 one can notice that the (new) indicators reflect different contents: the former makes more precise the role of team members through their entire participation in the hierarchical procedure, while the latter emphasizes the strength of the competition leading to the final ranking.  
 
 An additional contribution stems from the weight given by the ranking to the indicators, when displaying them as a function of the team ranks (Figs. \ref{Plot 3 D-L T-L d-h-m} - \ref{Plot 51 dh DL GdI TdF VaE}).

In fact, one observes the existence of clusters of teams, more explicitly in Fig. \ref{Plot 2 D-L T-L d-h-m}  as in many examples of socio-economics populations: a high, a medium and a low class of teams. Further investigation might be pursued through a recently proposed cluster stability indicator, Unit Relevance Index (URI).$^{\cite{CerquetiMattera2025}}$  

Let us recall that (alas) it is very difficult to move from one class to another, except through the introduction of external fields, - most likely money as the incentive. Inequalities and concentrations are inevitable but one can observe how extreme  cases are concerned. As examples, one has noticed that the  two best teams may interchange their rank according to the filtering, see Tables 3-5. Same for the worst teams. In both cases, very generally, the matter is very relevant for team management  and race strategies. However, in order to maintain some form of competition, one has not to neglect the “middle class" as the needed ballast.  Nevertheless, the most relevant features to be considered are those containing extreme values since they are expected to lead the search toward the main differences in team ranking, - and specifically here disentangling characteristics.

Whence one enters into the consideration about collusion, or competition-collaboration,$^{ \cite{caram1}}$ - which might change from stage to stage. Further  studies should concern whether such collusions can be observed in each stage, in other words how the teams move in the range around the extremum in the displayed curves,  particularly in the $\delta h$ plots. Clearly, the best teams have not much interest in bringing upward “middle class" teams, nor “middle class" teams bring upward  the “low class" teams. No need to say that  strategies allowing that a team could be better off by exerting a lower effort  maybe hidden in examining classical indicators,$^{\cite{CsatoEB2018,CsatoEJOR2023}}$  but could be highlighted through indicators based on differences between reasonably unbiased criteria. 

  Whence the study of “ranks resulting of strategies and skills" through the classical indices, but further  taking into account the rank as weight,  could add further values to the reliability of hierarchy findings, and promote attractive competition. 

\section{Conclusions} \label{conclusions}

Therefore, before optimizing strategies, one should be convinced of the validity
of  efficiency criteria. In the present report, one focuses on team ranking
when the valuation outcome depends much on the performance of a subset
of team members. It is argued and demonstrated through
  data found in cycling races. It is suggested that new indicators
be compared to classical ones. 
This leads to observe features, like inequalities, clusters,  “amount of competition", “race temperature", 
i.e., measures which provide  quantified meanings outlining
possible strategic goals   arising in many team competitions. It is
interesting to point that one can one observe management  strategies, through
indicator values comparison in a given race or when comparing races. Thus,
even within such case studies, quantitative measures suggest considerations
for further empirical modeling.

In conclusion, the    indicators  based on (i) the concept of difference between the distributions of data points, and (ii) on a probabilistic reasoning take into account  the team final competition measure in an information-like approach, - the Shannon entropy. The proposed methodology is practical, simple, and useful:  the study emphasizes that  the method is based on scientific rationality and logical principles. 
A desirable characteristic of inequality measures is the existence of a graphical analogy with the indicator. This can enhance interpretability and help communicate results to non-experts. The study  reported here above  presents such graphical characteristics allowing the valuation of team performance, - Figs. 3-6.  The new notion  of “crucial core rank", emphasising the main teams is well illustrated, and original. 

 In summary, the    analyses of the proposed
indicators
point to a few practical features.  For example,
 Table \ref{TableTdFGdVaEILn0M0_MandF} and Table \ref{TableTLALDL23Fteamtimescharacteristics}   allow to envisage race difficulties differences. The displayed values emphasize that the time distributions of $T_L$ and $A_L$ are similar for $GdI$ and $TdF$, but differ from those of  $VaE$:  moreover,  each skewness is positive indicating   wide distributions for the slowest teams.  Each  kurtosis   value  indicates    whether balanced competition occurs in races. The same deduction holds when observing the $\sigma$, much shorter in $TdF$, indicating a fiercer competition between the top teams.
 The  values of  $ \Delta_L^{(\#)}$, in Tables \ref{Table4TLALGdI23teams} - \ref{Table5TLALVaE23teams},   when  large,  indicate that there is team emphasis on a distributive role according to riders skills.  
  The  $\theta_L$ as a relative temperature index appears to be a measure of the distribution of riders kinetic energy at the end of a race (or stage). 
 An interesting output for race  organizers and team managers  arise from the ranking plots of  indicators; those allow to observe the  crucial race core  made of the  most competitive teams.
  
  In brief, the numerical values point to differences in team strategy and goals in a given race based on rider skill levels. 
  Thus, it seems admissible that the indicators are new useful measures, but surely need to be further examined and developed.

Notice that the definition of $T_L$, $A_L$, $\Delta_L$, and $\theta_L$ can be used not only for  measures after the final stage, but also for every stage; see Appendix B.  The same holds true for most of the defined and calculated coefficients here above, - but not in the first stage of the race, of course.  Extensions, e.g.,  to the team member finishing $place$ rather than their  $time$ can  be easily done.  
(This might be valuable when discussing colorful jerseys in such races.) The selection of team members including the leader(s) can be based on previous statistics relying on the indicators. This  enhances the  answer to the question about  designing and further optimizing strategies in a competitive environment.
 
Finally, 
 it might be interesting to further discuss the Stage and Race Temperature indicators, e.g.,  through illustrative examples in order to 
 appreciate the practical and theoretical  meanings of the "intensity" measures.
It can be suggested, as further research, that one constructs hypothetical races with only a very small number of teams  teams and choose sets of  $p^{(s)}_k$ values to see the behaviour of stage and race temperature indicators.  However, this 
 would demand
a different focus, - though surely  essential to practical coaching.   

The suggestion might clearly demand longer sets of investigations, in order to grasp a meaningful discussion on intensity sizes; this obviously demands several simulations covering several cases. 

Nevertheless,   races with only a few competing teams are not common.  
  World Rally Championship (WRC) competitions might be of interest, - but the teams ranking rules are very different from those in cyclist races.

Thus, last but not least, even though the paper contains an original contribution to a special type of sporting activity, the lessons learned can be of more general use for many other activities,  i.e., as long as   the outcomes  depend not only on the  team effort but also on the performance of a subset of members of the team.    Open questions remain on the collaboration-competition aspect of such races.

\newpage
 
\bigskip  
 \begin{centering}
     {\bf  Appendix    A:  
   \bigskip  

Considerations on the Kendall $\tau$ coefficient and the weighted Kemeny distance.}
\bigskip  
    
   \end{centering}

  \bigskip
  
How teams are ranked  might have    a substantial effect:  in sport, the prize money is  higher for the first teams than  for the others. Sometimes the last teams  face relegation and may loose  sponsors.  Thus, a swap in positions, due to different ruling, may be  crucial.  

   The difference in hierarchies, including the scattering of the results,  derived from  ranking rules,  can be classically measured through the  Kendall $\tau$ rank-rank correlation coefficient, or equivalently through  the Kemeny distance in terms of the number ($N$) of competing teams,  which reads  $K= N(N-1)(1-\tau)/4$,  when there is no ex aequo.$^{\cite{KemenySnell1962,HeiserD'Ambrosio2013}}$
However,  Can$^{\cite{Can2014}}$ and Csató$^{\cite{Csato2017}}$  point out that the Kendall $\tau$ coefficient does not take into account the precise position of dissimilarities when comparing  two linear ranking sets. In particular,  there is no discrimination   about the (teams) relative position in each list.

 In order to weight the position of discordant pairs, Csató has  proposed a   hyperbolic function: $w_C= 1/r$, $r\;\in\;[1,r_{M}-1]$, based on the lowest rank $r$   of an item of a discordant  pair.$^{\cite{Csato2017}}$  A smoother weight distribution, $w_A= \sqrt{1/r}$, has  also  been proposed.$^{\cite{EJOR}}$  Obviously, the classical Kendall coefficient $\tau$ corresponds to    choosing  a $w_K=1$ for permutations forcing one of the vectors to become identical to the other.          The procedure can be repeated, appropriately weighting the various swaps, whence  obtaining a “weighted Kemeny distance''  between pairs of ranks; they are called  
  $K_C$, $K_A$, and $K_K$ respectively.   
  
  For completeness, one can  observe the number of concordant $C$ and   discordant  pairs  $D$, obtain the “score" $S\equiv C-D$, thereafter the Kendall $\tau$ coefficient from $S/(C+D)$, - when there is no ex aequo in the considered vectors.  In the present  cases, $C+D \equiv M_L(M_L-1)/2$. The results are reported in Table \ref{TableKCAAppA}.
 
 Notice that practically, in order to compare the ranks of pairs of teams, it is first useful to organize the teams in alphabetical order, giving them the appropriate rank for a given indicator. In all studied cases in the main text, the ranking  chosen is that corresponding to  $T_L$, as in Tables     \ref{Table4TLALGdI23teams} - \ref{Table5TLALVaE23teams}. 

 Notice that  the   indicators distances are necessarily ordered: $K_K \ge K_A \ge K_C$, and  $A_L^{(\#)}$  is closer to $\Delta_L^{(\#)}$ than $T_L^{(\#)}$.

   \begin{table}\begin{center}  
\begin{tabular}{|c|c||c|c|c||c|c|c|c|c|c|c|} 
 \hline 
		& &	$D$	&	$S$	&	$\tau$	&	$K_K$	&	$K_A$	&	$K_C$	\\ \hline	\hline
. &$T_L^{(\#)};A_L^{(\#)}$ 	&	11	&	231	&	0.91304	&	11	&	4.0196	&	1.7011		\\
$GdI$ &$T_L^{(\#)};\Delta_L^{(\#)}$	&	118	&	17	&	0.067194	&	118	&	36.1731	&	12.6254		\\
. & $A_L^{(\#)} \Delta_L^{(\#)}$	&	107	&	39	&	0.15415	&	107	&	32.6804	&	11.2680		\\ \hline
. & $T_L^{(\#)};A_L^{(\#)}$ 	&	22	&	187	&	0.80952	&	22	&	6.5734	&	2.2169		\\
$TdF$ &$T_L^{(\#)};\Delta_L^{(\#)}$	&	110	&	11	&	0.047619	&	110	&	36.1289	&	13.3487		\\
. & $A_L^{(\#)} \Delta_L^{(\#)}$	&	88	&	60	&	0.23810	&	88	&	30.3798	&	11.7866		\\ \hline
. & $T_L^{(\#)};A_L^{(\#)}$ 	&	8	&	215	&	0.93074	&	8	&	3.5031	&	1.9793		\\
$VaE$& $T_L^{(\#)};\Delta_L^{(\#)}$	&	74	&	83	&	0.35931	&	74	&	25.6371	&	10.2048		\\
. & $A_L^{(\#)} \Delta_L^{(\#)}$	&	66	&	99	&	0.42857	&	66	&	22.8145	&	9.1486		\\ \hline
	\end{tabular}
 \caption{  Characteristics values leading to measures of the “Kemeny distance'' (Kendall $\tau$  without normalisation) between pairs of  indicators, for the 3 races examined in the main text: 
 number of discordant pairs $D$; score $S$; Kendall $\tau$ coefficient; Kemeny  distance $K_K (\equiv D)$ ($https://ncatlab.org/nlab/show/Kendall+tau+distance$). The weighted Kemeny distances, with either square root or mere (-1) hyperbolic weight distribution on permutations, are $K_A$ or $K_C$, respectively.}
 \label{TableKCAAppA}
 \end{center}
 \end{table}

 \newpage
 \bigskip  
 \begin{centering}
     {\bf  Appendix    B:  
   \bigskip  

Considerations for extensions to daily stages.}
\bigskip  
  
   \end{centering}
  \bigskip

One can sketch  how to specify the main text considerations  toward a daily ranking mechanism, within a multi-stage race.

Recall that $ t^{(\#)}_{i,s}$ is  the finishing time of  one of the 3 fastest riders ($i=1,2,3$)  of  team $(\#)$ for  stage $s$.
 The team final time $T_s^{(\#)}$ is
 \begin{equation}\label{Ateameqdaily}
  T_s^{(\#)}  = \Sigma_{i=1}^{3} \; \; t_{i,s}^{(\#)} \; .
\end{equation}  

For the first stage, $s=1$, of course, and  $T_1^{(\#)}   \equiv   A_1^{(\#)}$. After the second stage, one claims according to UCI rules that
 \begin{equation}\label{Ateameqs1s2}
  T_{2}^{(\#)}  \equiv T_1^{(\#)} + T_2^{(\#)} \;. 
\end{equation}  
After the 3rd stage,
 \begin{equation}\label{Ateameqs1s2s3}
  T_{3}^{(\#)}  \equiv T_1^{(\#)} + T_2^{(\#)} + T_3^{(\#)} \;, 
\end{equation}  
etc. 

Next, consider the adjusted team time after two stages $s=1$ and  $s=2$, i.e.,   $  A_{2} ^{(\#)}  $. 
This team adjusted time  is not equal to  $\Sigma_{i=1}^{3} \; \; t_{i,1}^{(\#)}
+ \Sigma_{i=1}^{3} \; \; t_{i,2}^{(\#)}$,  but is equal to the cumulative time of the best 3 ($j$) riders of the team $after$ the $s$-th stage, i.e.,  $= \Sigma_{j=1}^{3} \; \; t_{j,2}^{(\#)}$. Etc.

This seems to represent better the team time evolution and leads to avoid Cipollini-like effects.$^{\cite{ANORCipollini}}$
 
 \newpage
{\bf  Acknowledgements :}
 Thanks to  reviewers and editor for their patience and comments.
   
   Thanks to Prof. J. Miśkiewicz for much help on coding.

\bigskip  
{\bf   Data availability} : data is freely available,  see text. 

  \bigskip  
  {\bf   Funding} :   Work was partially supported by the project ‘’A better understanding of socio-economic systems using Quantitative Methods from Physics’’, funded by the European Union---Next generation EU and the Romanian Government under the National Recovery and Resilience Plan for Romania, contract no.760034/23.05.2023, code PNRR-C9-I8-CF 255/29.11.2022, through the Romanian Ministry of Research, Innovation and Digitalization, within Component 9, ‘’Investment I8’’. Moreover, P.K. acknowledges the support of ‘Digital Finance - Reaching New Frontiers’ (Horizon Marie Sklodowska-Curie Actions Industrial Doctoral Network), Ref. Number 101119635.
    
  \bigskip    
      {\bf  Disclosure Statement  on competing interest} :  Neither  relevant financial nor non-financial  competing interest has to be mentioned.
 \bigskip
 \\ \\


\clearpage

\begin{thebibliography}{99}
  
 \bibitem{Kulakowski20}
 1. Kulakowski K.   Understanding the analytic hierarchy process. Boca Raton, FL:  CRC Press;  2020.
 
 \bibitem{Fritzetal23}
2.  Fritz F, Moretti S, Staudacher J. Social ranking problems at the interplay between social choice theory and coalitional games. {\it Mathematics} 2023; 11(24): 4905. 

 \bibitem{AusloosGRRC24}  3. Ausloos M, Rotundo G,  Cerqueti R.  A theory of best choice selection through objective arguments grounded in linear response theory concepts. {\it  Physics} 2024; 6(2): 468-482. 

 \bibitem{Kossi17} 4. Kossi Y.  Tournois séquentiels et compétition pour la prime d’excellence scientifique. {\it  Rev Fr Econ} 2017; 32(4): 57-94. [in French]. Available from: $https://doi.org/10.3917/rfe.174.0057 $
 
  \bibitem{ref3academiachoice} 5. Sanz-Menéndez L,  Cruz-Castro L.  University academics’ preferences for hiring and promotion systems. {\it  Eur J High Educ} 2019; 9(2): 153-171.
  
   \bibitem{Jose} 6. Jose VRR, Nau RF.  Winkler RL.  Scoring rules, generalized entropy, and utility maximization. {\it  Oper Res} 2008; 56(5): 1146-1157.  
   
    \bibitem{CsatoIGTR2019} 7. Csató L.  Some impossibilities of ranking in generalized tournaments. {\it  Int Game Theory Rev} 2019; 21(01): 1940002. 
    
 \bibitem{ChebotarevShamis1998} 
 8. Chebotarev PYu., Shamis E. Characterizations of scoring methods for preference aggregation. {\it  Ann Oper Res} 1998; 80(9): 299–332.

  \bibitem{Fainmesseretal2009} 
 9. Fainmesser I, Fershtman C,  Gandal N.  A consistent weighted ranking scheme with an application to NCAA college football rankings.  {\it  J Sport Econ} 2009; 10(6): 582-600.
 
  \bibitem{GonzalezDiazetal2014}
  10.  González-Díaz J, Hendrickx R, Lohmann E.  Paired comparisons analysis:  an axiomatic approach to ranking methods. {\it  Soc Choice  Welf}  2014; 42(1): 139–169.

     \bibitem{Csato2017} 11. Csató L.  On the ranking of a Swiss system chess team tournament. {\it  Ann Oper Res} 2017; 254(1-2): 17–36.

\bibitem{Vazirietal2018}  
 
 12. Vaziri B, Dabadghao S, Yih Y, Morin TL. Properties of sports ranking methods. {\it  J Oper Res Soc}   2018; 69(5): 776–787.   
     
  \bibitem{Csato2023}  13. Csató L.  A comparative study of scoring systems by simulations. {\it  J Sport Econ} 2023; 24(4): 526–545.

  \bibitem{LeivaBertran2025}
      14. Leiva-Bertrán F.  Ranking in incomplete tournaments:  The generalized win percentage method, efficiency, and NCAA football. {\it  J Sport Econ} 2025; 26(1): 3–34.

   \bibitem{JianuAoR2023} 

15. Jianu I, Isaic-Maniu A, Brandas C, Cristescu MP, Bente C, Herteliu C. Testing Benford and universal laws on gambling and betting data in Romania. {\it  Ann Oper Res} 2023; 342(3): 1765–1779.

  \bibitem{CsatoEJOR2022}  
  
  16. Csató L. Quantifying incentive (in)compatibility:  A case study from sports. {\it  Eur J Oper Res}   2022; 302(2): 717–726. 
  
\bibitem{ausloos2017hintTdF} 
 
 17. Ausloos M. Hint of a Universal Law for the Financial  Gains of Competitive Sport Teams. The case of Tour de France cycle race. {\it  Front Phys} 2017; 5: 59.
 
\bibitem{ausloos2020concentrationTdF} 

 18. Ausloos M. Rank–size law, financial inequality indices and gain concentrations by cyclist teams. The case of a multiple stage bicycle race, like Tour de France. {\it  Physica A}   2020; 540(123161): 123161.   

\bibitem{Ausloos2023}  
 
 19. Ausloos M. Shannon entropy and Herfindahl-Hirschman index as team’s performance and competitive balance indicators in cyclist multi-stage races. {\it  Entropy}  2023; 25(6): 955.  

  \bibitem{EJOR}  
 
 20.  Ausloos M. Hierarchy selection:  New team ranking indicators for cyclist multi-stage races. {\it  Eur J Oper Res}   2024; 314(2): 807–816.  
 
 \bibitem{ANORCipollini}
 21.  Ausloos, M. Should one (be allowed to) replace the Cipollini’s?. {\it  Ann Oper Res} 2025;  in press. doi: 10.1007/s10479-024-06206-y.
 
 
\bibitem{Mostaertetal2021}
33. Mostaert M, Laureys F, Vansteenkiste P, Pion J, Deconinck FJ,  Lenoir M.  Discriminating performance profiles of cycling disciplines. {\it   Int J Sports Sci Coach}  2021; 16(1): 110-122.

\bibitem{PinedojauregiRomarate2025}
34. Pinedo-Jauregi A,  Romarate A.  Assessing ambient temperature measurements in road cycling races. {\it  Int J Sports Sci Coach}  2025;  20(2): 742-747. 

\bibitem{Smith2008}
35. Smith MF. Assessment influence on peak power output and road cycling performance prediction. {\it  Int J Sports Sci Coach} 2008; 3(2): 211-226. 

\bibitem{Stessesetal2024}
36. Stessens L, Gielen J, Meeusen R, Aerts JM. Physical performance estimation in practice:  A systematic review of advancements in performance prediction and modeling in cycling. {\it  Int J Sports Sci Coach} 2024; 19(5): 2222-2243. 

\bibitem{OGradyetal2023}
37. O’Grady MW, Worn R, Owens JO, O’Brie BJ.,  Talpey SW.  Race craft:  A qualitative exploration of the development, implementation and reflection of tactical decision making in road cycling. {\it  Int J Sports Sci Coach} 2023; 18(6): 2160-2170.



\bibitem{Sokolov2012}  

 22. Eliazar II, Sokolov IM. Measuring statistical evenness:  A panoramic overview. {\it  Physica A}   2012; 391(4): 1323–1353.   

\bibitem{Subramanian2012}  

 23. Subramanian N, Ramanathan R. A review of applications of Analytic Hierarchy Process in operations management. {\it  Int J Prod Econ}   2012; 138(2): 215–241.   

\bibitem{ZIDMA}  

 24. Dimitrova ZI, Ausloos M. Primacy analysis in the system of Bulgarian cities. {\it  Open Phys}   2015; 13(1): 218-225.   

\bibitem{JosaAguado2020}  
    
 25. Josa I, Aguado A. Measuring unidimensional inequality:  Practical framework for the choice of an appropriate measure. {\it  Soc Indic Res}   2020; 149(2): 541–570.   
    
\bibitem{Bednayetal25}  
 
  26. Bednay D, Fleiner B, Tasnádi A. An indifference result for social choice rules in large societies. {\it  Eur J Oper Res}   2025; 321(1): 208–213.   
 
  \bibitem{Atkinson1970}  
  
   27. Atkinson A.B. On the Measurement of Inequality. {\it  J Econ Theory} 1970; 2: 244-263.

\bibitem{Hirschman}  
 
  28. Hirschman AO. The paternity of an index.  {\it  Am Econ Rev} 1964; 54(5): 761-762.
  
\bibitem{Gini1921}  
 
 29. Gini C.   Measurement of Inequality of Incomes.  {\it  Econ J} 1921; 31(121): 124–125.
 
 \bibitem{Theil65}  

 30.   Theil H. The information approach to demand analysis. In:  {\it Advanced Studies in Theoretical and Applied Econometrics.} Dordrecht:  Springer Netherlands;  1992. pp. 627–651.
 
  \bibitem{MagesRohner2024}  
  
  31. 
 Mages T, Rohner C. Quantifying redundancies and synergies with measures of inequality. {\it PLoS One }2024;  19(11):  e0313281.
      
\bibitem{HallTideman67}  

 32. Hall M, Tideman N. Measures of concentration. {\it  J Am Stat Assoc}   1967; 62(317): 162–168.   


\bibitem{Shannon}  

38. Shannon CE.  A mathematical theory of communication. {\it  The Bell System Technical Journal} 1948; 27(3): 379-423. 

\bibitem{refTsallis}  
  
39. Tsallis C. Beyond Boltzmann–Gibbs–Shannon in physics and elsewhere. {\it  Entropy }   2019; 21(7): 696.   
    
\bibitem{Silvaetal2016}  

 40. Silva P, Duarte R, Esteves P, Travassos B, Vilar L. Application of entropy measures to analysis of performance in team sports. {\it  Int J Perform Anal Sport }  2016; 16(2): 753–768.   

\bibitem{refKozuki03}  
  
 41.  Kozuki N, Fuchikami N. Dynamical model of financial markets:  fluctuating ‘temperature’ causes intermittent behavior of price changes. {\it  Physica A }  2003; 329(1–2): 222–230.   
  
\bibitem{refLietal23}  

 42. Li H, Xiao Y, Polukarov M, Ventre C. Thermodynamic analysis of financial markets:  Measuring order book dynamics with Temperature and Entropy.  {\it  Entropy }   2023; 26(1): 24.   

\bibitem{BorooahManganfootball}  

  43. Borooah VK, Mangan J. Measuring competitive balance in sports using generalized entropy with an application to English premier league football. {\it  Appl Econ}   2012; 44(9): 1093–1102.   
 
\bibitem{Brezinaetal2016}  

 44. Brezina I, Pekár J, Čičková Z, Reiff M. Herfindahl–Hirschman index level of concentration values modification and analysis of their change. {\it  Cent Eur J Oper Res}   2016; 24(1): 49–72.   

 \bibitem{YigitTur2012}  

 45. Yiğit İ, Tür Ş. Relationship between diversification strategy applications and organizational performance according to Herfindahl index criteria. {\it  Procedia Soc Behav Sci }  2012; 58: 118–127.   

\bibitem{OladimejiUdosen2019}  

 46. Oladimeji MS, Udosen I. The effect of diversification strategy on organizational performance. {\it  J Compet }  2019; 11(4): 120–131.   

\bibitem{Handoyoetal2023}  

47. Handoyo S, Suharman H, Ghani EK, Soedarsono S. A business strategy, operational efficiency, ownership structure, and manufacturing performance:  The moderating role of market uncertainty and competition intensity and its implication on open innovation. {\it  J Open Innov}   2023; 9(2): 100039.  

\bibitem{OwenRyanW07}  
 
  48. Owen PD, Ryan M, Weatherston CR. Measuring competitive balance in professional team sports using the Herfindahl-Hirschman index. {\it  Rev Ind Organ}   2007; 31(4): 289–302.   
 
\bibitem{OwenOwen22}  

 49. Owen PD, Owen CA. Simulation evidence on Herfindahl-Hirschman measures of competitive balance in professional sports leagues. {\it  J Oper Res Soc}   2022; 73(2): 285–300.   


\bibitem{MarmaniKaur}  

51. Marmani S, Ficcadenti V, Kaur P, Dhesi G. Entropic analysis of votes expressed in Italian elections between 1948 and 2018. {\it  Entropy }   2020; 22(5): 523.  

\bibitem{QQ49.15.2307RCMAstatistical}  
 
 52. Cerqueti R, Ausloos M. Statistical assessment of regional wealth inequalities:  the Italian case. {\it  Qual Quant}   2015; 49(6): 2307–2323.   
 
 \bibitem{Lorenz1905}  

50. Lorenz, MO.  Methods of measuring the concentration of wealth. {\it   Publ Am Stat Assoc} 1905;  9(70): 209-219. 
 
\bibitem{FENSRszeszowAPPA129}  

 53.  Ausloos M, Cerqueti R. Studies on regional wealth inequalities:  The case of Italy. {\it  Acta Phys Pol A}   2016; 129(5): 959–964.  
  
\bibitem{Frosini2012}  

 54. Frosini BV. Approximation and decomposition of Gini, Pietra–Ricci and Theil inequality measures. {\it  Empir Econ}   2012; 43(1): 175–197.   

\bibitem{caram1}  

 55. Caram LF, Caiafa CF, Proto AN, Ausloos M. Dynamic peer-to-peer competition. {\it  Physica A}   2010; 389(13): 2628–2636.   

\bibitem{IJMPCFIFAMARCAGNV}  

 56 Ausloos M, Cloots R, Gadomski A, Vitanov NK. Ranking structures and rank–rank correlations of countries:  The FIFA and UEFA cases. {\it  Int J Mod Phys C}   2014; 25(11): 1450060.  

  \bibitem{CerquetiMattera2025}  
  
57. Cerqueti R, Mattera R.  Measuring unit relevance and stability in hierarchical spatio-temporal clustering. {\it  Spat Stat} 2025; 66: 100880. 

\bibitem{CsatoEB2018}  

58. Csató L. Was Zidane honest or well-informed? How UEFA barely avoided a serious scandal. {\it  Econ Bull } 2018; 38(1): 152-158.   

\bibitem{CsatoEJOR2023}  
 
  59. Csató L. How to avoid uncompetitive games? The importance of tie-breaking rules. {\it  Eur J Oper Res}   2023; 307(3): 1260–1269.   
 
 \bibitem{KemenySnell1962} 60.   Kemeny JG, Snell L. Mathematical models in the social sciences. Cambridge, MA:  MIT Press;  1962.
   
 \bibitem{HeiserD'Ambrosio2013} 61. Heiser WJ, D’Ambrosio A.  Clustering and prediction of rankings within a Kemeny distance framework. In: {\it   Lausen B, Van den Poel D, and Ultsch A (Editors)}. Algorithms from and for Nature and Life. Cham: Springer International Publishing;  2013, pp. 19–31.

\bibitem{Can2014} 62. Can B.  Weighted distances between preferences. {\it  J Math Econ} 2014; 51: 109–111.
 
    \end{thebibliography}
 \end{document}